\documentclass[traditabstract]{aa}
\usepackage{psfrag,graphicx}
\usepackage{txfonts}
\usepackage{natbib}
\usepackage{ulem}
\usepackage{color}
\usepackage{xspace}
\usepackage{supertabular}

\def \msol{M$_{\odot}$\xspace}
\def \pc{\,pc\xspace}
\def \kmps{\,km\,s$^{-1}$\xspace}
\def \cmcube{\,cm$^{-3}$\xspace}
\def \kmps{\,km\,s$^{-1}$\xspace}
\def \K{\,K\xspace}
\def \nhtot{$n_{\rm H}$\xspace}

\begin{document}

\bibliographystyle{aa}

%\title{The significance of gas-ice interplay in star-forming clouds}
\title{Interplay of gas and ice during cloud evolution}
\subtitle{}
\author{S. Hocuk \inst{1,2}
       \and
       S. Cazaux \inst{1}
       }
\institute{$^1$ Kapteyn Astronomical Institute, University of Groningen,
  P. O. Box 800, 9700 AV Groningen, Netherlands \\
  $^2$ Max-Planck-Instit\"{u}t f\"{u}r extraterrestrische Physik, Giessenbachstrasse 1, 85748 Garching, Germany\\
  \email{seyit@mpe.mpg.de, cazaux@astro.rug.nl}
}
%\titlerunning{gas-ice interplay in star-forming clouds}
\titlerunning{gas-ice interplay during cloud evolution}
\authorrunning{Hocuk, Cazaux}
\date{Received \today}

\abstract
{During the evolution of diffuse clouds to molecular clouds, gas-phase molecules freeze out on surfaces of small dust particles to form ices. On dust surfaces, water is the main constituent of the icy mantle in which a complex chemistry is taking place. We aim to study the formation pathways and the composition of the ices throughout the evolution of diffuse clouds. For this purpose, we used time-dependent rate equations to calculate the molecular abundances in the gas phase and on solid surfaces (onto dust grains). We fully considered the gas-dust interplay by including the details of freeze-out, chemical and thermal desorption, and the most important photo-processes on grain surfaces. The difference in binding energies of chemical species on bare and icy surfaces was also incorporated into our equations. Using the numerical code {\sc flash}, we performed a hydrodynamical simulation of a gravitationally bound diffuse cloud and followed its contraction. We find that while the dust grains are still bare, water formation is enhanced by grain surface chemistry that is subsequently released into the gas phase, enriching the molecular medium. The CO molecules, on the other hand, tend to gradually freeze out on bare grains. This causes CO to be well mixed and strongly present within the first ice layer. Once one monolayer of water ice has formed, the binding energy of the grain surface changes significantly, and an immediate and strong depletion of gas-phase water and CO molecules occurs. While hydrogenation converts solid CO into formaldehyde ($\rm H_2CO$) and methanol ($\rm CH_3OH$), water ice becomes the main constituent of the icy grains. Inside molecular clumps formaldehyde is more abundant than water and methanol in the gas phase, owing its presence in part to chemical desorption.}
%We keep track of the species abundances, their formation rates, and the constituents of the ices during this time. 
%This process is initially negligible, but becomes dominant toward to the formation of dense molecular clumps. 
%After 15.5 Myrs, water molecules freeze out more rapidly making water ice the main constituent of the icy grains. At this stage, dust grains no longer act as a catalyst for the gas phase and enrichment of gas-phase molecules from grain surfaces is halted until higher dust temperatures, $T_{\rm d} > 20$\,K, are reached. 
%We find that methanol ice is rapidly growing after 15 Myrs of cloud evolution. The final methanol coverage levels off around a composition of 43\% of the icy mantle. 

\maketitle

\section{Introduction}
\label{sec:introduction}
Dust grains play a vital role in the making and breaking of molecules in interstellar gas clouds because many chemical reactions proceed much faster on solid surfaces than reactions in the gas phase. As a result, dust can enhance molecular abundances in interstellar gas clouds by acting as a catalyst for the formation of complex molecules \citep[e.g., recent studies by][]{2012A&A...537A.102M, 2012A&A...541A..76L, 2012MNRAS.424.2961G}. However, dust grains can also lock up gas-phase molecules through freeze-out in cold environments \citep[e.g.,][]{1984MNRAS.209..955J, 2013A&A...560A..41L}. This duality creates an intricate relationship between gas and dust particles. Their involvement in chemical reactions affects the thermodynamic properties of molecular clouds and from this their whole evolution \citep{2001ApJ...557..736G, 2010A&A...522A..74C}. The kinetic temperature of a gas cloud is especially sensitive to changes in abundances of the dominant coolants, like CO \citep{2014MNRAS.438L..56H}. These findings highlight the importance of considering the formation of ices and depletion of gas-phase species in models and theories of cloud evolution and star formation.

Ices form on dust grains in the interstellar medium (ISM) during the evolution of interstellar clouds, the progenitors of star-forming regions. Diffuse clouds, where the dust-gas coupling and grain surface chemistry is still negligible, will evolve and undergo phase transitions to form molecular clouds, where freeze-out is effective, and eventually form dense clumps where dust-gas coupling becomes dominant. These are the critical phases that clouds undergo before a star finally forms and which, ultimately, determine the stellar masses at birth. During the evolutionary stages, gas-phase molecules are deposited on grain surfaces to form icy layers, thereby depleting the gaseous molecules. Eventually, the ices will grow thick mantels on dust grains and will remain there until the medium becomes warmer, allowing the species to evaporate back into the gas phase. Heating can be caused by radiation (UV photons, X-rays, or cosmic rays) penetrating the cloud and interacting with the gas, shock-waves injecting kinetic heating, or gravitational collapse where compressional heating ensues and radiation becomes trapped. Hot cores already revealed a glimpse of the rich organic chemistry locked up in the ice mantles \citep{2003ApJ...593L..51C}, which suddenly became visible as a result of evaporation into the gas phase by protostellar heating. 

Depletion of molecular species, such as CO, are also observed toward the formation of prestellar cores \citep{2006A&A...455..577T, 2013ApJ...775L...2L}, with depletion factors of up to 80 in dense clumps \citep{2012MNRAS.423.2342F}, indicating the presence of thick ice mantles. Observations of cold prestellar cores reported the lack of gas-phase H$_2$O, demonstrating a much more serious freeze-out than previously predicted \citep{2011PASP..123..138V}. The presence of frozen water was corroborated by the detection of strong water emission from shocks in protostellar environments \citep{2011PASP..123..138V, 2014MNRAS.440.1844S}. Recent observations of starburst galaxy M82 showed that CO$_2$ and H$_2$O ices are present for different physical conditions \citep{2013ApJ...773L..37Y}. This is intriguing because it is demanding for models to predict high abundances of CO$_2$. The pathway of forming CO$_2$ is still very uncertain \citep{2013A&A...554A..34I}, but the gas-phase route is widely accepted as inefficient. This leads to the notion that CO ice is not passive. %If anything, these results combined show us that we do not yet fully comprehend the formation of ices.

Observations of deeply embedded protostars show an anticorrelation between the abundance of CO (depleted) and methanol in the gas \citep{2010A&A...516A..57K}. This demonstrates that CO converts into methanol on the surface of dust grains and is subsequently released into the gas phase. The observed ortho-to-para ratio of water in the Orion Bar is also found to be inconsistent with only gas-phase water formation \citep{2013prpl.conf1S007C}, indicating that nonthermal processes must be at work on dust grains. An important mechanism to supply the gas phase with ice species is the nonthermal process of photodesorption from either UV or cosmic-ray-induced UV photons. Photons with energies of mostly around 8 eV can directly desorb CO molecules into the gas phase \citep{2011ApJ...739L..36F}. Photodesorption is considered as a possible important gas-phase supplier of CO \citep{2010A&A...522A.108M}, methanol and formaldehyde \citep{2013ApJ...764L..19Y, 2013A&A...560A..73G}, water \citep{2014MNRAS.440.2616K}, CO$_2$ \citep{2009A&A...496..281O, 2012ApJ...761...36B}, or O$_2$ \citep{2014MNRAS.437.3190Z}. Observations by \cite{2013A&A...558A..58Y} suggest, however, that photodesorption rates need to be increased by a factor of 2 to explain their abundances of O$_2$.

Recent experimental studies have unveiled another nonthermal mechanism, called chemical desorption, that directly converts species formed on dust surfaces into gas-phase species \citep{2007A&A...467.1103G, 2013NatSR...3E1338D}. 
This mechanism occurs for exothermic reactions, where the products cannot thermalize with the dust surface. These findings show that the formation of species through surface reactions does not just lock up species from the gas phase, but will also directly enrich the gas-phase medium and is therefore an integral part of interstellar cloud evolution.

In this work we track the formation of ices during the evolution of an interstellar gas cloud, starting from a diffuse, fully atomic stage until the formation of dense clumps. We observe the formation of the first ice layers on the surfaces of dust grains, report their compositions, and determine the distribution of ices around a dense clump as well as their formation rates. For this purpose, we developed a chemical network using rate equations that incorporates grain surface reactions on two different substrates, bare grain (no ices) surface, and water ice surface. We also consider the processes of chemical and photodesorption \citep{2010A&A...522A..74C, 2012MNRAS.421..768N, 2013NatSR...3E1338D} and the most recent reactions through quantum tunneling \citep{2012ApJ...749...67O, 2013A&A...559A..49M}. The paper is organized as follows: In Sect. \ref{sec:numericalmethod} we present the code that is used in this work and describe our initial conditions. In Sect. \ref{sec:analyticalmethod1} we explain the chemical processes on dust surfaces, describe each process separately, and present our equations. In Sect. \ref{sec:analyticalmethod2} we outline the important thermal processes and report the heating and cooling terms used in our calculations. In Sect. \ref{sec:results} we present our results on the composition of the ice layers, show the dominant species formation rates, and give the distribution of ices around a dense clump. We also discuss the implications of our results. In Sect. \ref{sec:conclusion} we conclude and discuss the caveats.

\section{Numerical method}
\label{sec:numericalmethod}
\subsection{Numerical code}
We performed the numerical simulations with the adaptive-mesh hydrodynamical code {\sc flash}, version 4.0 \citep{2000ApJS..131..273F, Dubey2009512}. Our work encompasses a broad range of physics, such as hydrodynamics, chemistry, thermodynamics (using time-dependent heating and cooling rates), turbulence, multispecies, gravity, and radiative transfer for UV. To solve the hydrodynamic equations, we applied the directionally split piecewise-parabolic method \cite[PPM;][]{1984JCoPh..54..174C}, which is well-suited to handling the type of calculations in this study. Our research captures the physics that act on small and large scales. These are the micro-sized scales required for the chemical reactions in the gas phase and on grain surfaces and the parsec-sized scales that are necessary to describe star formation processes. 

To track multiple fluids, we employed the multispecies unit that is able to follow each species with its own properties. We applied the consistent multifluid advection scheme \citep{1999A&A...342..179P} to prevent overshoots in the mass fractions as a result of the PPM advection. The Poisson equations were solved with the Multigrid solver in which gravity is coupled to the Euler equations through the momentum and energy equations. The physics modules are well-tested and were either provided by {\sc flash} or can be found in earlier works, for example, \cite{2010A&A...522A..24H, 2011A&A...536A..41H} and \cite{2014MNRAS.438L..56H}. A comprehensive chemistry and a thermodynamics module were created specifically for this work; they are explained in more detail in Sects. \ref{sec:analyticalmethod1} and \ref{sec:analyticalmethod2}.

\subsection{Initial conditions}
\begin{figure}[htb!]
\includegraphics[scale=0.102]{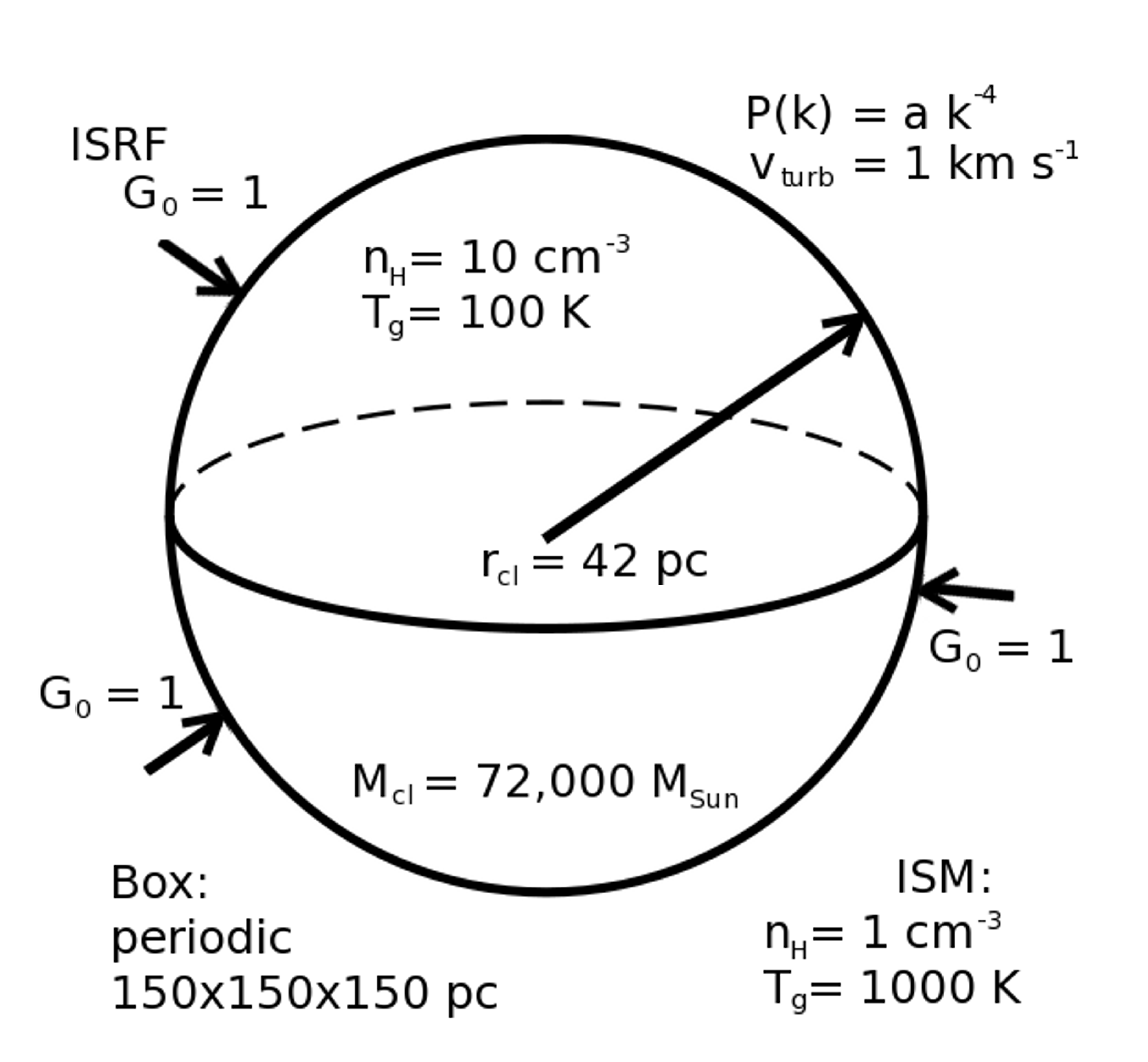}
\caption{Initial setup of the diffuse interstellar cloud.}
\label{fig:init}
\end{figure}

We created a gravitationally bound diffuse gas cloud in which all hydrogen is in atomic form. Our model cloud starts with a uniform number density of \nhtot = $10$\cmcube and an initial temperature of 100\K. \nhtot is defined as the total hydrogen nuclei number density, that is, \nhtot = $\rho/m_{\rm H}$. The interstellar environment including the cloud surface has a temperature of 1000\K and a number density of \nhtot = $1$\cmcube. We allowed for a smooth, hyperbolic transition for the values from the cloud edge to the surrounding ISM. To follow its chemistry and temperature during its evolution, we placed our model cloud in a 3D cubic box of size 150 pc$^3$. We applied periodic boundary conditions to our simulation domain. The spherical cloud has a radius of 42\pc and a total mass of 7.2$\times10^{4}$ \msol. 
A graphical display of the initial conditions is given in Fig. \ref{fig:init}.

The diffuse cloud was initiated with turbulent conditions that was representative of the ISM in the Milky Way, $\rm \sigma_{turb}$ = 1\kmps, with a power spectrum of $\rm P(k) \propto k^{-4}$ following the empirical laws for compressible fluids \citep{1981MNRAS.194..809L, 1999ApJ...522L.141M, 2004ApJ...615L..45H}. This scaling is also known as Burgers turbulence \citep{Burgers1939, 2007PhR...447....1B}. The turbulence is decaying and not driven. The cloud is also not supported by turbulence and will contract. The simulated cloud was placed in an environment with a background UV radiation flux of $G_0=1$ in terms of the Habing field \citep{1968BAN....19..421H}. This agrees with the ISM conditions of our Milky Way. The cloud center enjoys column densities of over 10$^{21}$cm$^{-3}$, which is equivalent to a visual extinction of $A_V \sim 0.5$. We do not consider magnetic fields in this work. 

We refined the computational grid that encloses the model cloud to a uniform resolution of 128$^3$ cells. This yields a spatial resolution of $1.17$\pc, which corresponds to a Jeans resolution of 44 grid cells per Jeans length for the initial state. The Jeans length is calculated as

\begin{equation}
\lambda_{\rm J} = \left( \frac {\pi \rm c_{ s}^2} {{\rm G} \rho} \right)^{\frac{1}{2}} ~\rm cm.
\label{eq:jeanslen}
\end{equation}

\noindent
Here $c_{s}$ is the sound speed, which, for an ideal gas, can be formulated as ${c_{\rm s}^{2}} = \gamma {P}/\rho = \gamma N_{\rm A} k_{\rm B} T / \mu$. In this, $N_{\rm A}$ and $k_{\rm B}$ are the Avogadro number and the Boltzmann constant, $\mu$ is the mean molecular mass, which is unity in the initial case, and the parameter $\gamma$ depends on the equation of state (EOS). For a polytropic EOS, the pressure scales as $\rm P \propto \rho^{\gamma}$ \citep{2000ApJ...538..115S}. The polytropic index $\gamma$ will be affected by the thermal balance, that is, heating and cooling, of the cloud and therefore is not a fixed quantity. The sound speed of the diffuse cloud if it is isothermal ($\gamma = 1$) is $\rm c_{s}=0.91$\kmps.

We did not increase grid resolution during the simulation nor allowed for star formation to occur. The spatial resolution will therefore not be as high as contemporary advanced numerical studies of star formation, but our goal is not to resolve the fine details of cloud fragmentation, turbulence, or the direct spatial conditions prior to star formation. Our focus lies on describing the chemical composition (gas + dust) and thermal balance of an evolving gas cloud.
%on performing and resolving the chemistry properly. More specifically, the formation of ices. 

%Chemical reactions on grain surfaces are generally much faster than reactions in the gas phase. However, because of the intricate connections in a large network among many species and, because of the interactions between the grain surface and the gas phase, an equilibrium solution is not acceptable.
Because the formation of ices can take more than 10$^4$ years \citep{2007ApJ...668..294C} and is dependent on the changing conditions within the cloud, an equilibrium solution is not acceptable. Within our time-dependent solver, we also did not allow for variations in species densities of over 5\% per iteration. To this end, we employed a very high time resolution with adaptive time-stepping that can extend to a time resolution of three days (0.01 yr) for a simulation that lasts over ten million years. In the most unfavorable case, it is necessary to iterate the chemistry routine more than 10,000 times within a hydrodynamical time step. The adaptive chemical time-stepping is handled within a subcycling loop of the hydrodynamical time step to avoid any unnecessary speed loss for the other routines. We let the cloud evolve until it reached 125\% of its theoretical free-fall time, where $t_{\rm ff} = \sqrt{ 3\pi / \rm 32G\rho}$. Given the initial conditions of the model cloud, this adds up to a final simulation time of $2.0\times 10^{7}$\,yr. %We stop our simulation before actual collapse sets in.

\section{Analytical method: time-dependent chemistry}
\label{sec:analyticalmethod1}
We performed time-dependent rate equations at each grid cell that include both gas and grain surface reactions to compute the chemical composition of the cloud. Our chemical model comprises 42 different species. Of these, 26 are in the gas phase and 16 are on dust grains. Our selection of species and their initial abundances (with respect to hydrogen atoms) is given in Table \ref{tab:species}. 
\begin{table}[htb!]
\begin{center}
\caption{List of species and their initial abundances ($n_{x_i}$/\nhtot).}
\begin{tabular}{lclc}
\hline
\hline
Species         & Initial abundance     & Species               & Initial abundance \\
\hline
H               & 1.0                   & HCO$^+$               & 0       \\
H$^{-}$         & 0                     & H$_{2}$CO             & 0       \\
H$^{+}$         & 0                     & CH$_{3}$O             & 0       \\
H$_{2}$         & 0                     & CH$_{3}$OH            & 0       \\
H$_2^{+}$       & 0                     & e$^{-}$               & 1.30$\times10^{-4}$     \\
H$_3^{+}$       & 0                     & $\bot$ H              & 0       \\
O               & 2.90$\times10^{-4}$   & $\bot$ H$_{\rm c}$    & 0       \\
O$^{-}$         & 0                     & $\bot$ H$_{2}$        & 0       \\
O$^{+}$         & 0                     & $\bot$ O              & 0       \\
O$_{2}$         & 0                     & $\bot$ O$_{2}$        & 0       \\
C               & 0                     & $\bot$ O$_{3}$        & 0       \\
C$^{-}$         & 0                     & $\bot$ OH             & 0       \\
C$^{+}$         & 1.30$\times10^{-4}$   & $\bot$ CO             & 0       \\
OH              & 0                     & $\bot$ CO$_{2}$       & 0       \\
OH$^{+}$        & 0                     & $\bot$ H$_{2}$O       & 0       \\
CO              & 0                     & $\bot$ HO$_{2}$       & 0       \\
CO$_{2}$        & 0                     & $\bot$ H$_{2}$O$_{2}$ & 0       \\
H$_{2}$O        & 0                     & $\bot$ HCO            & 0       \\
H$_{2}$O$^+$    & 0                     & $\bot$ H$_{2}$CO      & 0       \\
H$_{3}$O$^+$    & 0                     & $\bot$ CH$_{3}$O      & 0       \\
HCO             & 0                     & $\bot$ CH$_{3}$OH     & 0       \\
\hline
\end{tabular}
\label{tab:species}
\end{center}
\hspace{0.3cm} Note 1: The symbol $\bot$ denotes a bound/ice species.

\hspace{0.3cm} Note 2: $\rm \bot H_c$ is the chemically adsorbed counterpart of $\rm \bot H$.
\end{table}
There are 257 relevant reactions. Gas-phase reactions with the corresponding equations and their coefficients were obtained from the Kinetic Database for Astrochemistry \citep[\textit{KiDA};][]{2012ApJS..199...21W}. Surface reactions on dust grains were acquired from \cite{2010A&A...522A..74C} and comprise 89 reactions (see the appendix). %MORE REFERENCES?

\subsection{Chemistry solver}
Chemical reaction rates are solved using a fast and stable semi-implicit scheme, an improved scheme over the first-order backwards differencing (BDF) method developed by \cite{1997NewA....2..209A}. The derivation of this scheme is given below.

The general expression for implicitly solving ordinary differential rate equations is defined as

\begin{equation}
\frac{{\rm d}n_{x_i}}{{\rm d}t} = C_{x_i} - D_{x_i},
\label{eq:solver1}
\end{equation}
where $C_{x_i}$ is the creation and $D_{x_i}$ is the destruction rate of species $x_i$ at future, $t + {\rm d}t$, time step. The first-order integration of this differential equation is also known as the backward Euler method. It differs from a forward Euler method in which $C_{x_i}$ and $D_{x_i}$ would be based on current, known values. Knowing that $D_{x_i}$ implicitly depends on the species $x_i$ we can submit $D_{x_i} \equiv n_{x_i} D^{'}_{x_i}$, with $D^{'}_{x_i}$ the destruction rate coefficient \citep{2008A&A...490..521S}. Following this, the semi-implicit scheme \citep{1997NewA....2..209A} is obtained by

\begin{eqnarray}
\frac{{\rm d}n_{x_i}}{{\rm d}t} &=& \frac{n^{new}_{x_i} - n^{old}_{x_i}}{\Delta t} = C_{x_i} - n^{new}_{x_i} D^{'}_{x_i}, \nonumber
\\
n^{new}_{x_i} &=& \frac{n^{old}_{x_i} + C_{x_i} \Delta t}{1 + D^{'}_{x_i} \Delta t},
\label{eq:solver2}
\end{eqnarray}
where $old$ refers to the values at the current time step and $new$ points to the values at $t + \Delta t$, the future time step. Because the $new$ rates, $C_{x_i}$ and $D^{'}_{x_i}$, are also not known at the current time step, which we ideally wish to find, they can be approximated using current, $old$ values through iterating, applying a predictor-corrector method, or by using a mix between $old$ and $new$ values with the values that were previously evaluated. The latter does somewhat depend on the order in which the rates are calculated. 

We devised a second-order variant of the semi-implicit method (Eq. \ref{eq:solver2}) to solve our equations. We applied the trapezoidal rule to integrate our differential equation to gain higher precision. In its derived form, the second-order semi-implicit scheme is presented as %(SOSIS)

\begin{equation}
n^{j + 1}_{x_i} = \frac{n^{j}_{x_i} \left(1 - \frac{1}{2} D^{' j}_{x_i} \Delta t \right) + \frac{1}{2} \left( C^{j}_{x_i} + C^{j + 1}_{x_i} \right) \Delta t} {1 + \frac{1}{2} D^{' j + 1}_{x_i} \Delta t},
\label{eq:solver3}
\end{equation}
where we have replaced the time step descriptor by $j$ and $j + 1$ (previously, $old$ and $new$). This scheme is easily computed while still being an improvement over the explicit schemes and the first-order semi-implicit scheme. The order of error does not drop with respect to a second-order fully implicit method, that is, the local error is $O(h^3)$, the global error $O(h^2)$. Semi-implicit schemes are symplectic integrators in nature and yield far better results than standard Euler methods. We note, however, that for all semi-implicit BDF methods, one must ensure mass conservation, and they are not time reversible.

A higher order BDF method will be more accurate for solving the chemistry, as proven by \cite{2013MNRAS.434L..36B}, but will also come at a higher computational price. Employing a high time resolution with this scheme is, in our experience, sufficient to have good accuracy while still performing the calculations at an acceptable speed.

\subsection{Gas-phase chemistry}
Gas-phase reactions were acquired from the Kinetic Database for Astrochemistry \citep{2012ApJS..199...21W}. We considered every possible reaction (within the scope of the database) that involves our selection of species as given in Table \ref{tab:species}.

The 168 gas-phase reactions in our network include bimolecular reactions (e.g., $A + B \rightarrow C + D$), charge-exchange reactions (e.g, $A^+ + B \rightarrow A + B^+$), radiative associations (e.g., $A + B \rightarrow AB + photon$), associative detachment (e.g., $A^- + B \rightarrow AB + e^-$), electronic recombination and attachment (e.g., $AB^+ + e^- \rightarrow A + B$), ionization or dissociation of neutral species by UV photons, ionization or dissociation of species by direct collision with cosmic-ray particles or by secondary UV photons following H$_2$ excitation.

\subsection{Dust chemistry}
The solid-phase reactions or the reaction rate coefficients were gathered from \cite{2010A&A...522A..74C}. We have included several additional reactions in this work. The grain surface rate equations were simplified and set in easily accessible forms. To do this, we unified all the different reactions involving dust grains into five equation types. These are (A) adsorption of gas-phase species onto dust surfaces, (B) thermal desorption of ices, (C) two-body reactions on grain surfaces including chemical desorption, (D) cosmic-ray processes, and (E) photo-processes with UV photons that include photodissociation and photodesorption. The equations are explained in detail in the following five subsections.

\subsubsection{Adsorption onto dust grains}
\label{sec:accretion}
Species in the gas phase can be accreted onto grain surfaces. This depends on the motion of the gas species relative to the dust particle. Since the motions are dominated by thermal velocity, the adsorption rate depends on the square root of the gas temperature as $v_{th} = \sqrt{8 k_{\rm B} T_g/ \pi m}$. Once the gas species is in contact with the dust, there is a probability for it to stick on the surface of the grain. The sticking coefficient is calculated as

\begin{equation}
S(T) = \left( 1 + 0.4\left( \frac{T_g + T_d}{100} \right)^{0.5} + 0.2\frac{T_g}{100} + 0.08\left( \frac{T_g}{100} \right)^2 \right)^{-1},
\label{eq:sticking}
\end{equation}
where $T_g$ is the gas temperature and T$_d$ is the dust temperature \citep{1979ApJS...41..555H}. We note that this coefficient is based on H atoms. Using this, the adsorption rate can be formulated as,

\begin{eqnarray}
k_{ads} &=& n_d \sigma_d v_{x_i} ~~~\rm s^{-1}, \nonumber
\\
R_{ads} &=& n_{x_i} k_{ads} S(T) ~~~\rm cm^{-3} s^{-1},
\label{eq:eqchem1}
\end{eqnarray}
where $k_{ads}$ is the adsorption rate coefficient, $R_{ads}$ is the adsorption rate, $n_d$ is the number density of dust grains, $\sigma_d$ is the cross section of the grain, $v_{x_i}$ is the thermal velocity of species $x_i$, that is, $v_{x_i} = \sqrt{8 k_{\rm B} T_g/\pi m_{x_i}}$, with $m_{x_i}$ the mass in grams, and $n_{x_i}$ is the number density of the engaging species. In this equation, $n_d \sigma_d$ represents the total cross section of dust in cm$^{-1}$, which is obtained by integrating over the grain-size distribution. We adopted the grain-size distribution of \cite{1977ApJ...217..425M}, from here on MRN, with a value of $\alpha_{\rm MRN} = \langle n_d \sigma_d / n_{\rm H} \rangle_{\rm MRN} = 1\times10^{-21}$ cm$^{2}$. We chose this distribution rather than the one of \cite{2001ApJ...548..296W}, from here on WD, which has a total cross section larger by a factor three, that is, $\alpha_{\rm WD} = 3\times10^{-21}$ cm$^{2}$, because MRN did not include poly-cyclic aromatic hydrocarbons (PAHs). The freeze-out of species on PAHs to form ices is not known.

\subsubsection{Thermal desorption, evaporation}
\label{sec:evaporation}
After species are bound on grain surfaces, they can evaporate back into the gas. The evaporation rate depends exponentially on the dust temperature and on the binding energies of the species with the substrate. The binding energy of each species differs according to the type of substrate. We consider two possible substrates in this study, bare surfaces (assuming carbon substrate) and water ice substrate, since ices are mostly made of water. Species adsorbed on water ice substrate have in most cases binding energies higher than on bare dust or other ices \citep[e.g., see][]{2007ApJ...668..294C}. Species on top of CO, which can attain a significant coverage on dust, have binding energies that more closely resemble the binding energies of bare dust \citep[e.g.,][]{1988Icar...76..201S, 2014ApJ...781...16K}. Therefore, in this study we consider the binding energies on CO to be similar to those of bare dust.
% 
% CO, which can attain a significant coverage on dust, has a binding energy of $\sim$1000 K on a CO covered surface (960\,K \citealt{1988Icar...76..201S}; 990\,K \citealt{2014ApJ...781...16K}), whereas CO has a binding energy of around 1300\,K on water ice \citep{2012MNRAS.421..768N}. This means that the binding energy on a CO surface resembles the binding energy on a bare surface more closely than water ice. In this study we consider the binding energies on CO to be similar to that of bare dust.

We computed the fraction of the dust covered by (water) ice $\mathcal{F}_{ice}$ and bare $\mathcal{F}_{bare}$ to distinguish between the two. Together with the deposited amount of water ice, this fraction depends on the total number of possible attachable sites on grain surfaces per cubic cm of space, designated as $n_{d} n_{sites}$, which is defined as 

\begin{equation}
n_{d} n_{sites} = n_{d} \frac{4\pi r_d^2}{a_{\rm pp}^2} = n_{d}\sigma_d \frac{4}{a_{\rm pp}^2} \simeq 4.44\times10^{-6} n_{\rm H} ~~~{\rm cm^{-3} mly^{-1}},
\label{eq:ndns}
\end{equation}
where the radius of dust is given by $r_d$ and the typical separation between two adsorption sites on a grain surface is given by $a_{\rm pp}$, which we assume to be 3 \AA. A full monolayer (mly) is reached when all the possible sites on a grain surface are occupied by an atom or a molecule. To convert this from monolayers to number densities, one needs to multiply by $n_{d} n_{sites}$. %The total cross section, $n_{d}\sigma_d$ is, as stated before, $10^{-21} n_{\rm H}$ cm$^{-1}$ adopting an MRN distribution. 
For the dust-to-gas mass ratio $\epsilon_d$, intrinsic to the dust number density $n_{d}$, we assumed the typical value of $\epsilon_d=0.01$.

We obtained the fractions, $\mathcal{F}_{ice}$ and $\mathcal{F}_{bare}$, in the following manner: if the grain surface is covered by less than 1 mly of water ice,

\begin{equation}
\mathcal{F}_{ice} = \frac{n_{\bot H_2O}}{n_{d} n_{sites}}.
\label{eq:icefrac}
\end{equation}
When the grain is covered by more than 1 layer of water ice, $\mathcal{F}_{ice} = 1$. The bare fraction of the dust is obtained by

\begin{equation}
\mathcal{F}_{bare} = 1 - \mathcal{F}_{ice}.
\label{eq:barefrac}
\end{equation}

We note that at this stage we assumed that the bound species are homogeneously distributed and neglected the cases where stratified layers of species can form. This assumption can lead to an over- or underestimation of reaction rates because the species abundances on different substrates can differ. We expect, however, that since the situation $\mathcal{F}_{bare} \sim \mathcal{F}_{ice}$ is not very common, the over- or underestimation will be marginal. With these definitions, we can formulate the evaporation rate as follows:

\begin{eqnarray}
k_{evap} &=& \nu_0 \left( \mathcal{F}_{bare} \exp\left(-\frac{E_{bare,i}}{T_{d}}\right) + \mathcal{F}_{ice} \exp\left(-\frac{E_{ice,i}}{T_{d}}\right)\right) ~~~\rm s^{-1}, \nonumber
\\
R_{evap} &=& n_{x_i} k_{evap} ~~~\rm cm^{-3} s^{-1},
\label{eq:eqchem2}
\end{eqnarray}
where $k_{evap}$ is the evaporation rate coefficient, $R_{evap}$ is the evaporation rate, $\nu_0$ is the oscillation frequency, which is typically $10^{12}$ s$^{-1}$ for physisorbed species, %REFERENCE??
$E_{bare,i}$ and $E_{ice,i}$ are the binding energies of species $x_i$ on bare grains and ices. The species specific binding energies can be found in Table \ref{tab:bindingenergies}.

\begin{table}[htb!]
\begin{center}
\caption{Binding energies for the substrates bare grain and water ice.}
\begin{tabular}{lll|lll}
\hline
\hline
& \multicolumn{2}{c}{Substrate} & & \multicolumn{2}{c}{Substrate} \\
%\cline{2-3}
%\cline{5-6}
Species                 & Bare (K)      & Ice (K)       & Species               & Bare (K)        & Ice (K) \\
%Species        & E$_{\rm bare}$ (K)    & E$_{\rm Ice}$ (K) & Species & E$_{\rm bare}$ (K)    & E$_{\rm Ice}$ (K) \\
\hline
$\bot$ H                & 500$^{~eg}$   & 650$^{~d}$    & $\bot$ CO$_2$          & 2300$^{~h}$   & 2300$^{~h}$ \\
$\bot$ H$_{\rm c}$      & 10000$^{~c}$  & 10000$^{~c}$  & $\bot$ H$_2$O          & 4800$^{~ai}$  & 4800$^{~ai}$ \\
$\bot$ H$_2$            & 300$^{~j}$    & 300$^{~j}$    & $\bot$ HO$_2$          & 4000$^{~j}$   & 4300$^{~d}$ \\
$\bot$ O                & 1700$^{~j}$   & 1700$^{~j}$   & $\bot$ H$_2$O$_2$      & 6000$^{~j}$   & 5000$^{~d}$ \\
$\bot$ O$_2$            & 1250$^{~j}$   & 900$^{~dh}$   & $\bot$ HCO             & 1100$^{~m}$   & 3100$^{~l}$ \\
$\bot$ O$_3$            & 2100$^{~j}$   & 1800$^{~d}$   & $\bot$ H$_2$CO & 1100$^{~m}$   & 3100$^{~i}$ \\
$\bot$ OH               & 1360$^{~d}$   & 3500$^{~d}$   & $\bot$ CH$_3$O & 1100$^{~m}$   & 3100$^{~l}$ \\
$\bot$ CO               & 1100$^{~bg}$  & 1300$^{~hk}$  & $\bot$ CH$_3$OH        & 1100$^{~m}$   & 3100$^{~l}$ \\
\hline
\end{tabular}
\label{tab:bindingenergies}
\end{center}
$^{a}$ \cite{1988Icar...76..201S} \\
$^{b}$ \cite{2003Ap&SS.285..633C} \\
$^{c}$ \cite{2004ApJ...604..222C} \\
%$^{c}$ \cite{2004MNRAS.354.1133C} \\
$^{d}$ \cite{2007ApJ...668..294C} \\
$^{e}$ \cite{2010A&A...522A..74C} \\
$^{f}$ \cite{2010A&A...522A.108M} \\
$^{g}$ \cite{2011ApJ...735...15G} \\
$^{h}$ \cite{2012MNRAS.421..768N} \\
$^{i}$ \cite{2012A&A...543A...5N} \\
$^{j}$ \cite{2013NatSR...3E1338D}, with updated $\bot$O binding energies. \\
$^{k}$ \cite{2014ApJ...781...16K} \\
$^{l}$ we relate these to H$_2$CO$^{i}$ binding energies. \\
$^{m}$ we relate these to CO$^{bg}$ binding energies.
\end{table}

\subsubsection{Two-body reactions on dust grains}
\label{sec:twobody}
While species are attached to grain surfaces, they can move around by thermal diffusion and meet other species with which they can react to form new molecules. The mobility of the species depends on the oscillation frequency $\nu_0$. We only consider physisorbed species in this work, except for neutral hydrogen, where we also take chemisorption into account. In addition to the mobility, the reaction rate depends on the specific binding energy with the substrate and on the dust temperature. When two species meet, they can immediately react if there is no (or little) reaction barrier. If the reaction barrier is high, however, it might be crossed by tunneling. The probability of overcoming the reaction barrier by tunneling is given by

\begin{equation}
P_{\rm reac} = \exp\left( -(2a/\hslash)\sqrt{2 \, m_{\rm red} \, k_{\rm B} \, E_a} \, \right),
\label{eq:tunneling}
\end{equation}
where $a=1$\,\AA, %WHYYYYYYY?
$m_{\rm red}$ is the reduced mass of the two engaging species, that is, $m_{\rm red}=(m_i \times m_j) / (m_i + m_j)$, $h$ is the Planck constant, and $E_a$ is the energy of the barrier needed for the reaction to occur. The probability is $P_{\rm reac}=1$ if there is no barrier for the reaction to take place. We did not consider the reaction diffusion competition \citep{2011ApJ...735...15G}. If this were included, the tunneling probabilities would become much higher.%, on the order of unity, if $P_{reac}$ is greater than the migration probability, $P_{migrate}$, and proportional to $P_{migrate}^{-1}$ when $P_{reac} \ll P_{migrate}$. The latter is arguable.

For CO hydrogenation reactions that have a barrier, that is, $\bot$H+$\bot$CO and $\bot$H+$\bot$H$_2$CO, we deduced an `effective' tunneling barrier energy from the reaction rates obtained by \cite{2004ACP.....4.1461A} and \cite{2009A&A...505..629F} at low ($\sim$10\,K) temperatures. The activation barriers we calculated are 600\,K and 400\,K, respectively. The height of these barriers are verified from recent experimental results (Minissale et al. priv. comm.). More precise measurements are being performed. The used activation barriers can be found in the appendix.

When a reaction occurs, the product either remains on the surface or immediately desorbs into the gas phase because of the exothermicity of the reaction. The probabilities of desorption are given by $\delta_{\rm bare}$ and $\delta_{\rm ice}$ for our two substrates. The fraction that remains on the surface will be $1 - \delta$.
Non-exothermic reactions that do not desorb are by definition mutiplied by 1.
With this information, we formulate the two-body reaction rate on grain surfaces as

\begin{eqnarray}
k_{2body} &=& \nu_0 \mathcal{F}_{bare} \left( \exp\left(-\frac{2}{3}\frac{E_{bare,i}}{T_d}\right) + \exp\left(-\frac{2}{3}\frac{E_{bare,j}}{T_d}\right)\right)(1\textrm{-}\delta_{\rm bare}) \nonumber
\\
& + & \nu_0 \mathcal{F}_{ice~~} \left( \exp\left(-\frac{2}{3}\frac{E_{ice,i~~}}{T_{d}}\right) + \exp\left(-\frac{2}{3}\frac{E_{ice,j~~}}{T_{d}}\right)\right)(1\textrm{-}\delta_{\rm ice}), \nonumber %~\rm s^{-1}
\\
R_{2body} &=& \frac{n_{x_i} n_{x_j}}{n_{d} n_{sites}} P_{\rm reac} k_{2body} ~~~\rm cm^{-3} s^{-1},
\label{eq:eqchem3}
\end{eqnarray}
where $k_{2body}$ is the two-body rate coefficient, $R_{2body}$ is the two-body reaction rate. The exponent in this equation represents the diffusion of species on the surface, and we assume that diffusion occurs with a barrier of two-thirds of the binding energy (67\% \citealt{2013NatSR...3E1338D}, 40\% \citealt{2003Ap&SS.285..633C}, 90\% \citealt{2007ApJ...658L..37B}). The desorption rate is obtained from the complement of $1-\delta$ of the rate coefficient. 

The desorption probabilities $\delta_{\rm bare}$ for the exothermic reactions $\rm \bot H+\bot O \rightarrow OH$, $\rm \bot H+\bot OH \rightarrow H_2O$, $\rm \bot O+\bot O \rightarrow O_2$ 
% \begin{eqnarray}
% \rm \bot H &+& \rm \bot H \rightarrow \rm H_2 \nonumber \\
% \rm \bot H &+& \rm \bot O \rightarrow \rm OH \nonumber \\
% \rm \bot H &+& \rm \bot OH \rightarrow \rm H_2O \nonumber \\
% \rm \bot O &+& \rm \bot O \rightarrow \rm O_2 \nonumber
% \end{eqnarray}
were adopted from \cite{2013NatSR...3E1338D} and are $\delta_{\rm bare}=0.5, 0.9, 0.6$. For the hydrogenation reactions $\rm \bot HCO + \bot H \rightarrow H_2CO$ and $\rm \bot CH_3O + \bot H \rightarrow CH_3OH$, we assumed a desorption fraction of $\delta_{\rm bare} = 5\%$ and for $\rm \bot H_2CO + \bot H \rightarrow CH_3O$ $\delta_{\rm bare} = 50\%$. %to account for the fact that the reactions are exothermic, but that the products are heavier (first estimates from Minissale et al. in prep.). The weaker bound hydrogenated species HCO and CH$_3$O have 10 times higher desorption probabilities. 
%These values will be provided with more accuracy from future laboratory experiments (Minissale et al. in prep.). 
In our next paper, we will report the different chemical desorption yields on different types of surfaces, estimated by laboratory experiments (Cazaux et al. in prep.). For the desorption probabilities on icy substrates $\delta_{\rm ice}$, we considered that chemical desorption is much weaker than on bare surfaces and assumed that $\delta_{\rm ice} = \delta_{\rm bare}/5$ \citep[deduced from][]{2013NatSR...3E1338D}. We did not take into account that chemical desorption has a dependence on surface coverage \citep{2014JChPh.141a4304M}.

\subsubsection{Cosmic-ray processes on grain surfaces}
\label{sec:crs}
Cosmic-ray reaction rates on grain surfaces are assumed to be the same as the rates found in the gas phase. We included several of the cosmic-ray reactions for surface bound species. Cosmic-ray processes are usually inefficient destruction mechanisms, but can dominate the destruction rates deep inside the cloud. These reaction rates depend on the cosmic-ray ionization rate per H$_2$ molecule, which we adopted as $\zeta_{\rm H_2} = 5\times10^{-17}$ s$^{-1}$ \citep{2007ApJ...671.1736I, 2011A&A...536A..41H, 2012A&A...537A.138C}. The cosmic-ray reaction rate is formulated as

\begin{eqnarray}
k_{CR} &=& z_{x_i} \zeta_{\rm H_2} ~~~\rm s^{-1},  \nonumber
\\
R_{CR} &=& n_{x_i} k_{CR} ~~~\rm cm^{-3} s^{-1},
\label{eq:eqchem4}
\end{eqnarray}
where $k_{CR}$ is the cosmic-ray rate coefficient, $R_{CR}$ is the cosmic-ray reaction rate, and $z_{x_i}$ the cosmic-ray ionization rate factor that is subject to the ionizing element (see \textit{KiDA} database).

Cosmic-ray-induced UV (CRUV) photons were also considered within the same equation (Eq. \ref{eq:eqchem4}). In this case, $z_{x_i}$ is replaced by $z_{\rm CRUV}$, the UV photon generation rate per cosmic-ray ionization. UV photons from cosmic rays do not suffer from radiation attenuation as normal UV photons do. Hence, the lack of dependence on optical depth in this formula. 
%This is because cosmic rays can easily penetrate every part of the cloud, with only a weak dependence on density and column, %REFERENCCEEE PADELIS?
%while the UV photons are generated on-site of the cosmic ray collision. This allows such UV photons to be homogeneously distributed accross the cloud.

\subsubsection{Photo-processes on grain surfaces}
\label{sec:uvreactions}
When UV photons arrive on a dust particle, they can interact with the adsorbed species and either photodissociate or photoevaporate them. We used the same formula for both types of photo-processes. Photoreactions scale linearly with the local radiation flux (erg cm$^{-2}$ s$^{-1}$). The radiation field strength is necessarily a function of extinction, which is given by $\xi_{x_i} A_V$, where $\xi_{x_i}$ is the extinction factor that is contingent on the relevant species. We obtain $A_V$ by dividing the column density $N_{\rm H}$ over the scaling factor, that is, $A_V = N_{\rm H}$/2.21$\times$10$^{21}$ mag \citep{2009MNRAS.400.2050G}. The column densities were computed by integrating the density from the simulation boundaries to each point. Simply, this is $N_{\rm H} = \Sigma_i \, n_{\rm H} \, ds_i$, with $ds_i$ being the path length of the smallest resolution element in which the density remains constant. For this purpose, we constructed a ray-tracing algorithm using 14 equally weighted rays with long characteristics (traveling from inside to outside). The ray separation is set by our highest resolution. We assumed an isotropic UV radiation field with a flux of $1 \, G_0$, where $G_0 \equiv 1.6\times10^{-3}$ erg cm$^{-2}$ s$^{-1}$. %This corresponds to a Draine field of $\chi = 2.74\times10^{-3}$ erg cm$^{-2}$ s$^{-1}$.
The general photo-process rate equations are defined as
%UVVVV ENERGY BAND 6-13.6 eV?

\begin{eqnarray}
k_{phot} &=& \alpha_{x_i} e^{-\xi_{x_i} A_V} ~~~\rm s^{-1},
\\ \nonumber
R_{phot} &=& n_{x_i} f_{\rm ss} k_{phot} F_{\rm UV} ~~~\rm cm^{-3} s^{-1}.
\label{eq:eqchem5}
\end{eqnarray}
where $k_{phot}$ is the photo-process rate coefficient, $R_{phot}$ is the photo-process reaction rate, $\alpha_{x_i}$ is the unattenuated rate coefficient, $f_{\rm ss}$ is the self-shielding factor, and $F_{\rm UV}$ is the UV flux in units of $1.71 \, G_0$, that is, $G_0 = 1$ gives $F_{\rm UV} = 0.58$. The factor 1.71 arises from the conversion from the often used Draine field \citep{1978ApJS...36..595D} to the Habing field for the far ultraviolet (FUV) intensity. We used the same $\alpha_{x_i}$, $\xi_{x_i}$, and $f_{\rm ss}$ for the gas phase as we did for the surface reactions.

When UV photons arrive on an icy surface with multiple layers, we only allowed the first two layers to be penetrated by UV photons. As shown by \cite{2006JChPh.124f4715A}, \cite{2010JChPh.132r4510A}, and \cite{2010A&A...522A.108M}, only the uppermost few layers contribute to photodesorption. Photodissociation does seem to occur deeper into the ice, but trapping and recombination of species tend to dominate \citep{2008A&A...491..907A}. This means that the highest number density that the photons can see is $n_{x_i} = {\rm min}(n_{x_i},2n_{d}n_{sites})$. This restriction is also enforced for reactions with CRUV photons, as given in Sect. \ref{sec:crs}.

Self-shielding, denoted as $f_{\rm ss}$, was taken into account for H$_2$ and CO molecules. These molecules can shield the medium against photo-processes on grain surfaces as well as for species in the gas phase. The self-shielding factor for H$_2$ was obtained from \cite{1996ApJ...468..269D}, Eq. 37, which is formulated as %STEPHS paper would be nice here2007ApJ...671.1736I

\begin{equation}
f_{\rm ss} = \frac{0.965} {(1 + x/b_5)^2} + \frac{0.035} {(1 + x)^{0.5}}  \times {\rm exp}\left[-8.5\times 10^{-4} (1 + x)^{0.5}\right],
\label{eq:selfshield}
\end{equation}
where $x \equiv N_{\rm H_2} / 5\times 10^{14}$ cm$^{-2}$, $b_5 \equiv b/10^5$ cm\,s$^{-1}$, and $b$ is the line broadening of H$_2$ lines, which we take as 3 km\,s$^{-1}$. This factor is only a function of H$_2$ column density $N_{\rm H_2}$. We used the column density algorithm to compute the H$_2$ column for this purpose. For CO molecules, self-shielding was achieved by incorporating the self-shielding tables from \cite{2009A&A...503..323V} into our code. Given an H$_2$ column and a CO column, the factor $f_{\rm ss}$ was acquired from the tables. For all other species we took $f_{\rm ss} = 1$.

Photodesorption is only implemented here for $\bot$CO molecules. Since CO molecules in the gas phase can strongly affect the thermal balance of collapsing molecular clouds, as we have shown in an earlier work \citep{2014MNRAS.438L..56H}, detailed processes were taken into account to correctly determine gas-phase abundances of CO. This encompasses both the normal and the CRUV photons. Photodesorption is not expected to be the dominant destruction mechanism of $\bot$CO, but is a route to desorb surface-bound CO molecules directly into the gas phase. The two implemented reactions are
\begin{eqnarray}
\bot \rm CO & + & {\rm UV \, photon} \rightarrow \rm CO, \nonumber
\\
\bot \rm CO & + & {\rm CRUV \, photon} \rightarrow \rm CO. \nonumber
\end{eqnarray}
For these two photo-chemical reactions $\alpha_{x_i}$ of Eq. \ref{eq:eqchem5} represents the number of CO photodesorptions per second per unit radiation flux \citep{2009ApJ...690.1497H, 2012A&A...537A.138C}. This variable depends on the photodesorption yield, which was experimentally obtained and adopted by us as $Y_{\rm CO} = 1.0\times 10^{-2}$ \citep{2011ApJ...739L..36F}. With respect to previous values \citep{2009A&A...496..281O}, these yields are higher by about a factor 4. In addition, only the first two monolayers were assumed to be penetrable by UV photons for these rates.
%Munoz Caro et al 2010 find even higher yileds, i.e. about 5.4e-2 at 8K ice.

\section{Analytical method: Thermodynamics}
\label{sec:analyticalmethod2}
To address the thermodynamics of the gas cloud, we calculated time-dependent heating and cooling rates that complemented the chemistry calculations. In this way, we obtained the gas and dust temperatures by solving the thermal balance.

\subsection{Heating and cooling}
We included the most prominent heating processes that are relevant to our work. The non-equilibrium heating processes include
\begin{itemize}
\item photoelectric heating, 
\item H$_2$ photodissociation heating, 
\item H$_2$ collisional de-excitation heating, 
\item cosmic-ray heating,
\item gas-grain collisional heating (when $\rm T_{gas} < T_{dust}$).
\end{itemize}
Compressional heating and shock heating are by default taken into account through the hydrodynamics, the EOS, and the shock-detection routines of our code, {\sc flash}. 

The cooling of the gas is ensured by several different types of non-equilibrium processes. These processes are 
\begin{itemize}
\item electron recombination with PAHs cooling, 
\item electron impact with H (i.e., Ly-$\alpha$) cooling, 
\item metastable transition of [OI]-630 nm cooling, 
\item fine-structure line cooling of [OI]-63 $\mu$m and [CII]-158 $\mu$m,
\item molecular ro-vibrational cooling by H$_2$, CO, OH, and H$_2$O,  
\item and gas-grain collisional cooling (when $\rm T_{gas} > T_{dust}$).
\end{itemize}
Cooling by adiabatic expansion was, again, handled by the standard EOS routines of {\sc flash}. The heating and cooling functions and their rates are described in \cite{2014MNRAS.438L..56H}.

\subsection{Dust temperature}
\begin{figure}[htb!]
\includegraphics[scale=0.51]{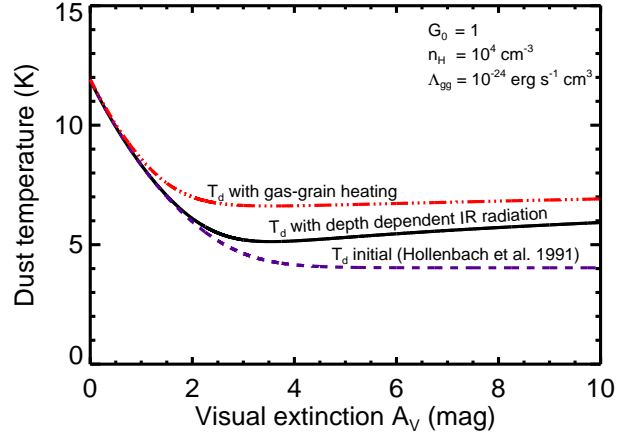}
\caption{Dust temperature estimated by including additional heating sources. The purple dashed line displays the initial estimate of the dust temperature as given by \cite{1991ApJ...377..192H}. The black solid line is the dust temperature that takes into account the depth dependence of infrared radiation by using a simple scaling with opacity. The red dot-dashed line also considers the gas-grain heating for which the variables $\Lambda_{\rm gg}$ and \nhtot are fixed here to serve in this example.}
\label{fig:tdust}
\end{figure}

The dust temperature is a crucial parameter that not only influences the gas temperature through the heating and the cooling rates, but also affects the chemical reaction rates. This eventually drives the formation and build-up of ices. We follow \cite{1991ApJ...377..192H}, also mentioned in \cite{2012A&A...540A.101L}, but with some adaptations of our own. The initial estimate of the dust temperature, $T_{d,i}$, which incorporates the attenuated incident UV radiation field, the cosmic microwave background temperature $T_{\rm CMB} = 2.725$ K, and the infrared dust emission, is defined as

\begin{eqnarray}
T_{d,i}^5 &=& 8.9\times10^{-11} \nu_0 G_0 \, {\rm exp}(-1.8 A_V) + T_{\rm CMB}^5 + 3.4\times10^{-2} \times \nonumber
\\
& & \left[ 0.42 - {\rm ln}(3.5\times10^{-2} \tau_{100} T_0) \right] \tau_{100} T_0^6,
\label{eq:tdust}
\end{eqnarray}
where the adopted value $\nu_0 = 3\times10^{15}$ s$^{-1}$ represents the most efficient absorbing frequency over the visual and UV wavelengths, $\tau_{100}$ is the emission optical depth at 100 $\mu$m, and $T_0$ is the equilibrium dust temperature at the cloud surface due to unattenuated incident FUV field alone \citep{1991ApJ...377..192H}. $T_0$ in this case equates to $T_0 = 12.17 G_0^{1/5}$ K. If it is assumed that the incident FUV flux equals the outgoing flux of dust radiation from $T_0$, then $\tau_{100} = 2.7\times10^3 G_0 T_0^{-5}$ \citep{1991ApJ...377..192H}. Knowing $T_0$ fixes $\tau_{100}$ to a value of 0.001. This makes it independent of optical depth. We display this by plotting the dust temperature as a function of visual extinction in Fig. \ref{fig:tdust}, which also shows other dust temperature calculations.

To accommodate for the depth dependence of the infrared emission, we calculated $\tau_{100}$ in a different fashion. Since $A_V = 2.5 {\rm log}_{10}(e) \, \tau_V \simeq 1.086 \, \tau_V$, where $\tau_V$ is the opacity at optical (predominantly 550 nm) wavelengths, and because

\begin{equation}
\tau_{\lambda} = \tau_{100} \left( \frac{100 \, \mu{\rm m}}{\lambda} \right)^{\beta_{\rm sed}},
\label{eq:taunu}
\end{equation}
with $\lambda$ being the wavelength and $\beta_{\rm sed}$ the sub-mm slope of the spectral energy distribution known as the spectral emissivity index, we can rewrite $\tau_{100}$ as a function of $A_V$ in the form

\begin{equation}
\tau_{100} = \tau_{550 \, \rm nm} \left( \frac{550 \, {\rm nm}}{100 \, \mu{\rm m}} \right)^{1.3} = 9.4\times10^{-4} A_V.
%4.60\times10^{-3} A_V.
\label{eq:newtau}
\end{equation}
For $\beta_{\rm sed}$ we take the value of 1.3, which gives the same $\tau_{100}$ at $A_V \simeq 1$ as \cite{1991ApJ...377..192H} advocated. A spectral emissivity index between $\beta_{\rm sed}=1-2$ is typically found for the Milky Way \citep{2012A&A...538A.137M, 2012A&A...541A..19A}. However, we now have a higher value of $\tau_{100}$ at higher $A_V$ to account for the depth dependence.

We also considered the heating of dust grains by the gas through the gas-grain collisional heat exchange. Since dust grains have a larger heat capacity, the heating of dust grains will be considerably weaker than the cooling of the gas, but this process might still slightly increase the dust temperature. Assuming that there is a temperature equilibrium in which the equilibrium timescale is much shorter than a free-fall time, the energy balance to reach a stable dust temperature is given by

\begin{equation}
\Lambda_{\rm gg} = \Gamma_{\rm rad},
\label{eq:ebalance}
\end{equation}
where $\Lambda_{\rm gg}$ is the gas-grain collisional heating and $\Gamma_{\rm rad}$ is the radiative losses due to blackbody radiation. Note that these are the losses from raising the dust temperature to a higher value by gas-grain collisional heat exchange than the equilibrium temperature given in Eq. \ref{eq:tdust}. The losses can be described as

\begin{equation}
\Gamma_{\rm rad} = 4\sigma_{\rm SB} \left( T_d^4 - T_{d,i}^4 \right) \rho_d \kappa_P ~~~\rm erg ~cm^{-3} ~s^{-1},
\label{eq:ebalance2}
\end{equation}
with $\sigma_{\rm SB}=5.67\times10^{-5}$ erg cm$^{-2}$ s$^{-1}$ K$^{-4}$ the Stefan-Boltzmann constant, $\rho_d$ the mass density of dust, and $\kappa_P$ the Planck mean opacity. Here, we also make use of our initial estimate of the dust temperature $T_{d,i}$ and expect $T_{d,i}$ to be at a stable equilibrium dust temperature when there is no additional heating source. We adopt the relation 

\begin{equation}
\kappa_P = 4\times10^{-4} T_d^2 ~~~\rm cm^2 ~g^{-1}
\label{eq:ebalance3}
\end{equation}
as presented by \cite{2000ApJ...534..809O} for a typical molecular cloud composition \citep{1994ApJ...421..615P} and for $T_d \lesssim 50$ K, but we did not consider Rosseland mean opacity in our solution \citep{2007A&A...475...37S, 2011ApJ...729L...3D}. Combining Eqs. \ref{eq:ebalance} and \ref{eq:ebalance2}, we obtain

\begin{equation}
T_d^4 = \frac{\Lambda_{\rm gg}}{4\sigma_{\rm SB} \rho_d \kappa_P} + T_{d,i}^4.
\label{eq:ebalance4}
\end{equation}
Replacing $\kappa_P$ with Eq. \ref{eq:ebalance3} and by using $\rho_d = \epsilon_d \rho$ for the dust mass density, with $\epsilon_d = 0.01$, we can formulate

\begin{equation}
T_d^6 = \frac{\Lambda_{\rm gg}(T_{d*})}{9.072\times10^{-10} \rho} + T_{d,i}^4 T_{d*}^2,
\label{eq:ebalance5}
\end{equation}
where $T_{d*}$ is a prediction for the real dust temperature. This so-called sextic equation is a transcendental equation and unsolvable analytically for $T_{d*} = T_d$, but we can solve it numerically by way of iteration while taking the first-order estimate of the dust temperature as $T_{d*} = T_{d,i}$. 

In the end, we have a slightly higher, more accurate dust temperature than our initial estimate that increases with density as a result of better gas-grain coupling, and with optical depth for infrared radiation. To highlight the difference in dust temperature by our adjustments, the three different dust temperatures are displayed together in Fig. \ref{fig:tdust}.

\section{Results}
\label{sec:results}
\begin{figure}[htb!]
\includegraphics[scale=0.44]{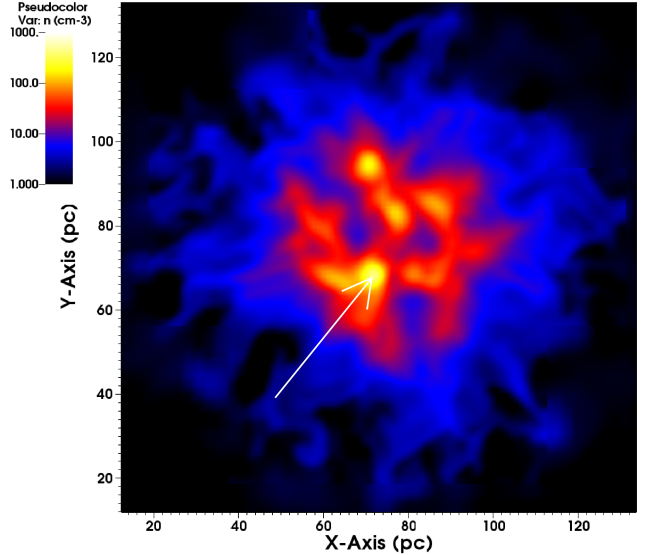}%0.44 0.40
\caption{Clumpy structure of a translucent cloud. Density slice along the $Z$-axis of the evolving cloud at t = 1.27$\times 10^{7}$ yr after initiation. The white arrow indicates the dense clump that is used for the results in this work.}
\label{fig:core}
\end{figure}

When we started our simulation, the turbulence created a filamentary structure of the gas inside the cloud. It slowly evolved into a more clumpy structure with dense clumps that were gravitationally unstable. Ten million years after the initial phases, we chose the densest part of our cloud, a collapsing clump of radius $r_{\rm clump} = 1 \, \rm pc$ to follow its evolution. The density of our clump grew from \nhtot $\sim$ 10$^2$ to $10^{4.5}$ \cmcube during this time. The clump was located close to the cloud center. In Fig. \ref{fig:core} we display this clump after nearly 13 Myr of simulation. 

To investigate how the chemistry adapts, we recorded the species abundances, their formation rates, and the thermodynamic quantities in time and in space. The results we report as a function of time are based on the mean value of the cells located inside the chosen clump. We also present plots of the clump conditions pertaining to a region of space at a fixed time. In this case, we show the growth of ice layers as a function of visual extinction and present maps in Cartesian coordinates. To also expose the outer regions, we then examine a greater radial distance, with $r_{\rm map} = 5 \, \rm pc$.

\subsection{Evolution of the clump}
\begin{figure}[htb!]
\includegraphics[scale=0.49]{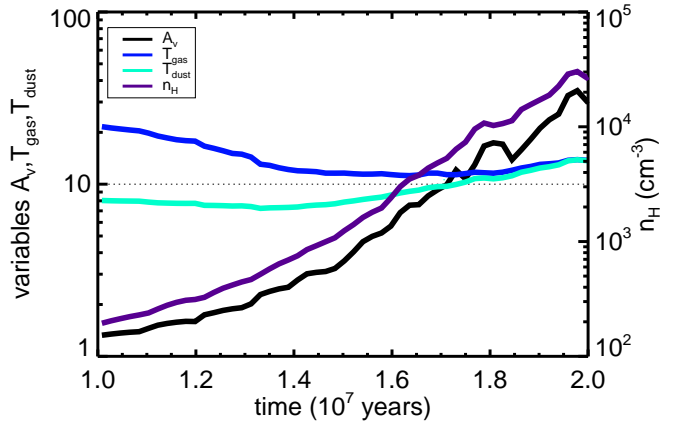}
\caption{Time evolution of clump parameters. The variables $A_V$ (black), T$_{\rm gas}$ (blue), T$_{\rm dust}$ (light blue), and \nhtot (purple) are plotted as a function of dynamical time.}
\label{fig:clumpvar}
\end{figure}

The density of the clump after 10 million years of cloud evolution is \nhtot = 200 \cmcube, corresponding to an $A_V$ of near unity. This increases to a 3$\times10^4$ \cmcube after another 10 million years, see Fig. \ref{fig:clumpvar}.
Both the gas temperature and the dust temperature of the clump within the same time interval (10 - 20 Myr) initially decrease as the clump becomes denser, and they rise as a result of compressional heating after rapid collapse sets in. At densities of \nhtot = $2\times10^4$ \cmcube and above, gas-dust coupling allows the two physical states to enjoy the same temperature, with $\rm T \gtrsim 12 K$. This occurs around an $A_V$ of 20. The optical depth is high because the clump is deeply embedded inside the cloud, close to its center. The collapse of our clump is not delayed, nor does it fragment during the course of the simulation.

\subsection{Time evolution of species abundances}
\begin{figure}[htb!]
\includegraphics[scale=0.51]{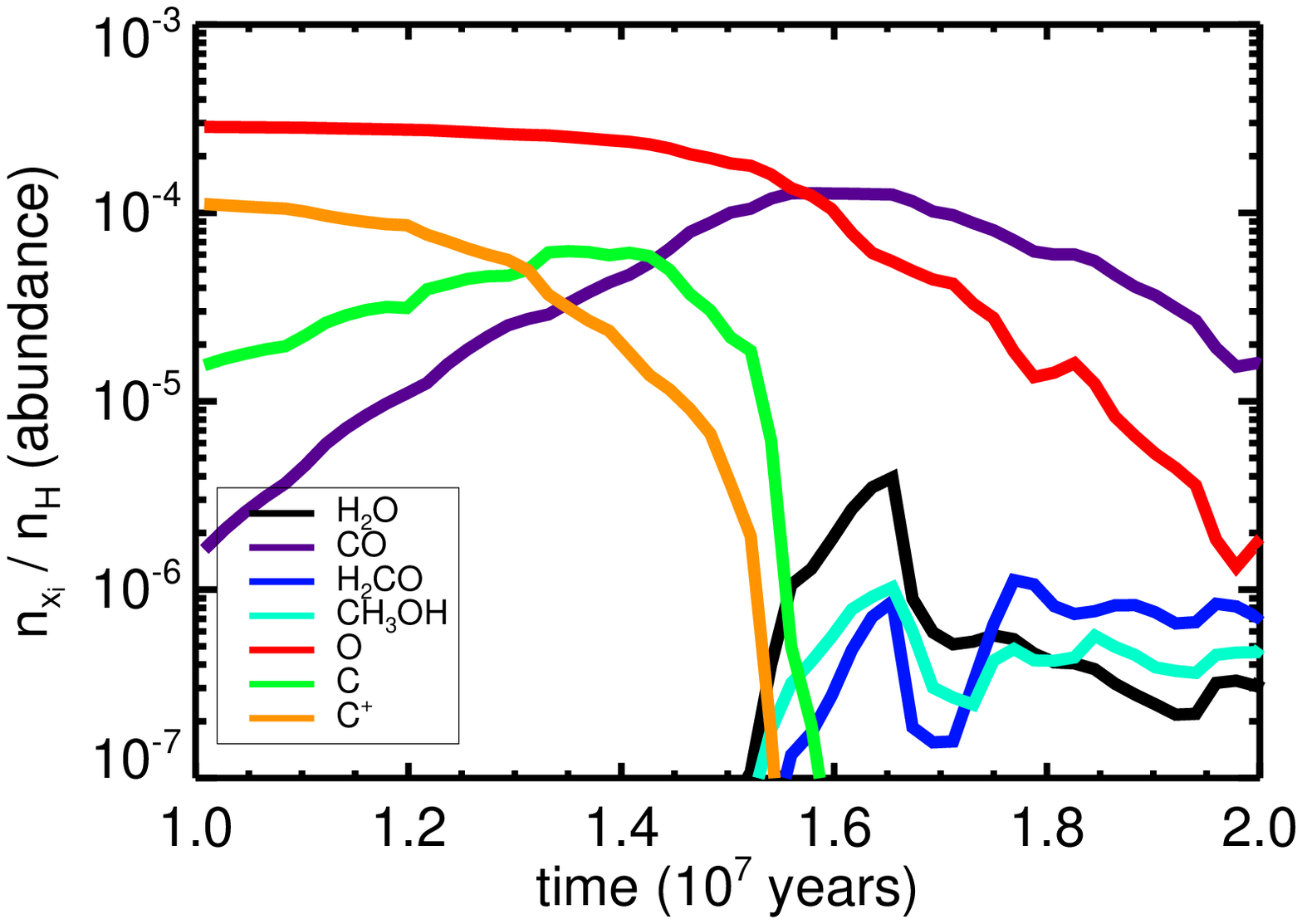} \\
\includegraphics[scale=0.51]{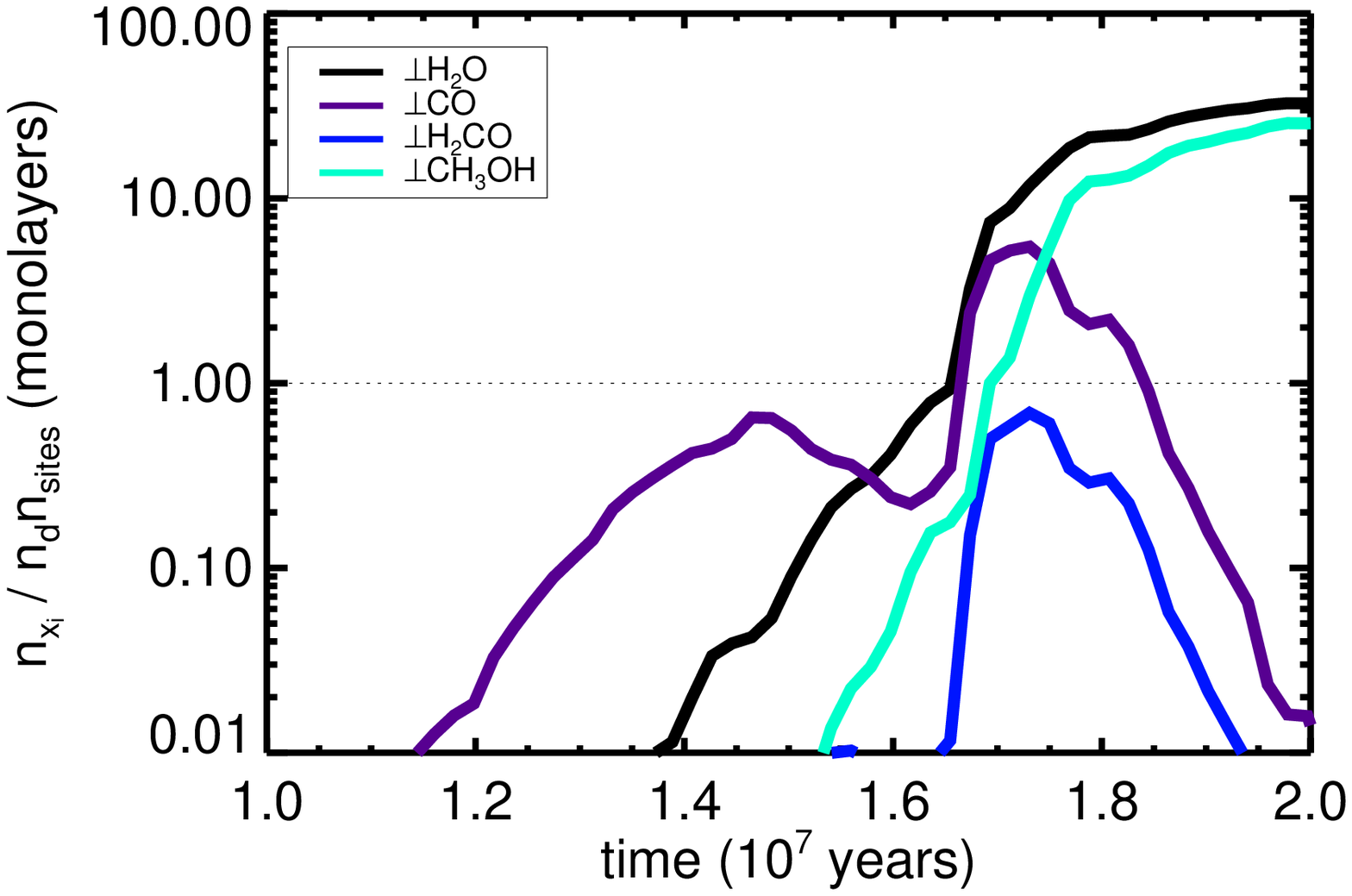} 
\caption{Species abundances varying with time. The top panel displays the abundance of gas-phase species as a function of time. The bottom panel shows the growth of ice layers on grain surfaces. Each species is represented by a different color.}
\label{fig:icegrowth}
\end{figure}

We plot in Fig. \ref{fig:icegrowth} the time evolution of the species abundances in the gas phase and on grain surfaces. These are the results of the abundances within our 1 pc clump. 
%
%
% These are the results of the abundances within our 1 pc clump. 

\noindent \\
\textit{Gas-phase species abundances}\\
In the upper panel of Fig. \ref{fig:icegrowth}, where we plot the gas-phase species, the amount of ionized carbon decreases with passing time and as the cloud becomes denser, while in the meantime the neutral carbon abundance increases. Carbon is readily converted into CO when the cloud enters a translucent stage at the clump density of \nhtot = $10^3$ \cmcube and at $A_V=3$. CO peaks at t = 1.6$\times10^7$ yr, with a peak abundance of $n_{\rm CO}/n_{\rm H} = 1.28\times10^{-4}$, which amounts to 99\% of all the carbon. This occurs at a density of \nhtot = $3\times10^3$ \cmcube while the visual extinction has reached $A_V = 5$. The only significant way to deplete CO from the gas phase beyond this point is through CO freeze-out on grain surfaces. We can see this happening by the decrease in CO abundance after t = 1.65$\times10^7$ yr. A full layer of water ice has covered the dust surface at this point. See Fig. \ref{fig:icegrowth} bottom panel where the black (water) line crosses the dotted line. The CO abundance drops to $\sim 2\times10^{-5}$ near the end of our simulation. We perceive that it continues to drop to $\sim 2\times10^{-6}$ at t = 2.2$\times10^7$ yr, which extends beyond the plot range. %whereafter the descend slows down around
Oxygen is also increasingly depleted from the gas phase when the dust is enveloped by a mantle of water ice. Neutral oxygen has a steeper decline than CO, with an abundance of $n_{\rm O}/n_{\rm H} = 1.5\times10^{-6}$ at t = 2$\times10^7$ yr. This means that oxygen is also being locked up in something other than CO, which we know to be mostly water ice. 

Within our densest clump, we never reach a situation with a C/O abundance ratio of unity. A C/O ratio above unity might lead to interesting carbon chemistry in the gas phase and is attained by models using varying local interstellar background radiation fields, $G_0 = 1-1000$ \citep{2009ApJ...690.1497H}. We remind that we use $G_0 = 1$. The ratio we obtain always remains below unity and consistently below $0.01$ above t = 1.55$\times10^7$ yr, when more than half of the dust surface is covered.

The gas-phase abundances of water (black), methanol (light blue), and formaldehyde (dark blue) rise above a value of $10^{-7}$ after 15 Myr (Fig. \ref{fig:icegrowth}), but remain relatively constant with abundances that, in the same order, linger around 3.2, 4.7, $7.5\times10^{-7}$ at the end of our simulation. The H$_2$CO/CH$_3$OH ratio at that moment is 1.6, which is a consequence of the high chemical desorption of formaldehyde. The abundances of species inside the clump after 20 Myr of cloud evolution are listed in Table \ref{tab:species2}.
\begin{table}[htb!]
\begin{center}
\caption{Abundances of species at the end of the simulation ($n_{x_i}$/\nhtot).}
\begin{tabular}{lclc}
\hline
\hline
Species         & Final abundance       & Species               & Final abundance \\
\hline
H               & 2.1$\times10^{-4}$    & HCO$^+$               & 8.4$\times10^{-11}$     \\
H$^{-}$         & 9.4$\times10^{-13}$   & H$_{2}$CO             & 7.5$\times10^{-7}$      \\
H$^{+ }$                & 7.7$\times10^{-10}$   & CH$_{3}$O             & $<10^{-20}$     \\
H$_{2}$         & 0.49                  & CH$_{3}$OH            & 4.7$\times10^{-7}$      \\
H$_2^{+}$       & 6.5$\times10^{-11}$   & e$^{-}$               & 6.2$\times10^{-7}$      \\
H$_3^{+}$       & 2.9$\times10^{-9}$    & $\bot$ H              & 2.0$\times10^{-10}$     \\
O               & 1.5$\times10^{-6}$    & $\bot$ H$_{\rm c}$    & 2.2$\times10^{-6}$      \\
O$^{-}$         & 1.2$\times10^{-16}$   & $\bot$ H$_{2}$        & 1.1$\times10^{-6}$      \\
O$^{+}$         & 6.8$\times10^{-16}$   & $\bot$ O              & 1.1$\times10^{-12}$     \\
O$_{2}$         & 3.4$\times10^{-6}$    & $\bot$ O$_{2}$        & 2.1$\times10^{-10}$     \\
C               & 2.6$\times10^{-10}$   & $\bot$ O$_{3}$        & 5.8$\times10^{-13}$     \\
C$^{-}$         & 1.1$\times10^{-16}$   & $\bot$ OH             & 2.8$\times10^{-12}$     \\
C$^{+}$         & 1.7$\times10^{-13}$   & $\bot$ CO             & 6.2$\times10^{-8}$      \\
OH              & 3.0$\times10^{-7}$    & $\bot$ CO$_{2}$       & 2.4$\times10^{-8}$      \\
OH$^{+}$        & 1.1$\times10^{-12}$   & $\bot$ H$_{2}$O       & 1.5$\times10^{-4}$      \\
CO              & 1.5$\times10^{-5}$    & $\bot$ HO$_{2}$       & 2.5$\times10^{-12}$     \\
CO$_{2}$        & 5.1$\times10^{-9}$    & $\bot$ H$_{2}$O$_{2}$ & 5.7$\times10^{-13}$     \\
H$_{2}$O        & 3.2$\times10^{-7}$    & $\bot$ HCO            & 8.6$\times10^{-12}$     \\
H$_{2}$O$^+$    & 3.3$\times10^{-12}$   & $\bot$ H$_{2}$CO      & 8.5$\times10^{-9}$      \\
H$_{3}$O$^+$    & 4.7$\times10^{-11}$   & $\bot$ CH$_{3}$O      & 7.7$\times10^{-12}$     \\
HCO             & 8.5$\times10^{-10}$   & $\bot$ CH$_{3}$OH     & 1.1$\times10^{-4}$      \\
\hline
\end{tabular}
\label{tab:species2}
\end{center}
\hspace{0.3cm} Note 1: The symbol $\bot$ denotes a bound/ice species.

\hspace{0.3cm} Note 2: $\rm \bot H_c$ is the chemically adsorbed counterpart of $\rm \bot H$.
\end{table}

We reach a high amount of methanol ice at the end of our simulation. This is partially because we did not incorporate species larger than methanol in our network. In addition to this, we have not considered diffusion through tunneling for oxygen atoms \citep{2013PhRvL.111e3201M}, which should lead to a lower methanol and a higher CO$_2$ abundance. 

\noindent \\
\textit{Ice species abundances}\\
The lower panel of Fig. \ref{fig:icegrowth} displays the abundance of frozen species in terms of ice layers covering the surface of the dust, which are given in units of monolayers along the $Y$-axis. Dust grains within the clump grow thicker ice mantles as a function of time, which starts to level off around $\rm t = $ $1.8\times10^7$ yr. The maximum number of ice layers is mainly limited by the total surface area and the availability of the oxygen atoms, $n_{\rm O}/n_{\rm H} = 2.9\times 10^{-4}$ (Table \ref{tab:species}). We reach a total of 59 ice layers at the end of our simulation. This means that eventually 97\% of the oxygen freezes out on dust and that only 3\% resides in the gas phase locked in CO. The majority of ices are in the form of $\bot$H$_2$O, $\bot$CO, $\bot$H$_2$CO (formaldehyde), or $\bot$CH$_3$OH (methanol). 

When the cloud is still in a diffuse stage early on in cloud evolution, $\bot$CO is the predominant surface-bound species, with frozen water only second to $\bot$CO. This means that the initial ice mantle is well mixed. This contradicts the idea that the first ice layer should consist of water ice alone. The $\bot$CO/$\bot$H$_2$O ratio is about unity when half the dust surface is covered. This ratio decreases quickly over time due to the repeated hydrogenation of $\bot$CO to eventually form formaldehyde and methanol. The amount of $\bot$CO steeply decreases after the first ice layer has formed, while water ice grows more rapidly with increasing density. CO and water each dominate the surface coverage at different epochs, but water ice always dominates the ice composition when more than one layer of ice covers the dust. The dominance of solid CO on surfaces lasts for about 3 Myr in our simulation, which is only a fraction of a cloud dynamical time, whereafter water ice becomes the dominant ice species. We expect that frozen CO can only be observed mixed with ices. %with $\Delta t_{\rm CO} \lesssim 5 \times 10^6$ yr

As stated earlier, the first mly of ice is formed at t = 1.65$\times10^7$ yr after we start our simulation. More importantly than time, the first ice layer is formed when the cloud is dense and cold enough, at which point it enters a molecular stage. Before that, we find that solid species start to grow rapidly, that is, beyond 0.1 mly, above a density of \nhtot = 360 \cmcube, with n$_{\rm H_2}$/n$_{\rm HI}$ = 2, and below a gas temperature of T$_g$ = 16\,K. The cloud environment where these frozen species form has an $A_V \geq 1.8$ with a dust temperature of 7.5 K. Water ice becomes the main constituent of the icy surface when the gas density rises above \nhtot $= 2.0\times10^3$ and the gas temperature falls below 11.5\,K. The visual extinction of our clump now reaches upto 5.2. This also marks the turnover point where gas-phase CO starts to become depleted and the gas and dust temperatures start rising. More than two-thirds of the dust surface is covered by ice at this time. This in turn shifts the adsorption energies of the species from bare surface to icy surface binding energies, see Table \ref{tab:bindingenergies}. In most cases, the binding energies increase when the substrate becomes water ice. These changes affect the thermal evaporation and the reaction rates on dust surfaces, as can be seen by Eqs. \ref{eq:eqchem2} and \ref{eq:eqchem3}. Where the former one causes a build-up of ices by inhibiting evaporation, the latter slows down the formation of species by reducing the mobility. Moreover, the chemical desorption probabilities also drop by a factor of 5 when the dust is covered by ice, adding to the ice build-up. The combined effects work in favor of constructing more ices. This is shown by the sharp rise in the curves in Fig. \ref{fig:icegrowth} around the point where one mly of ice has formed.

\subsection{Time evolution of species formation rates}
\begin{figure*}[htb!]
\includegraphics[scale=0.51]{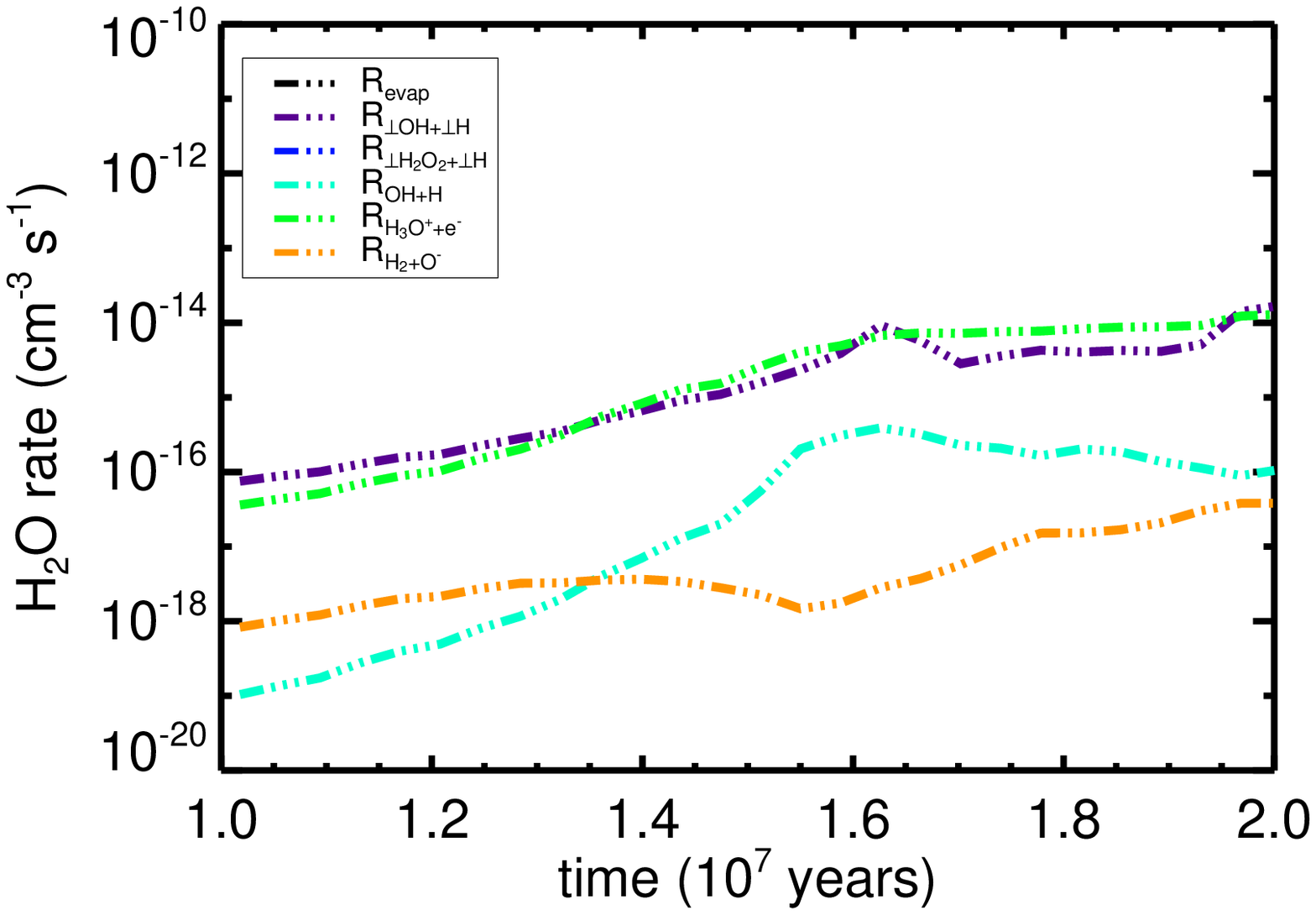}
\includegraphics[scale=0.51]{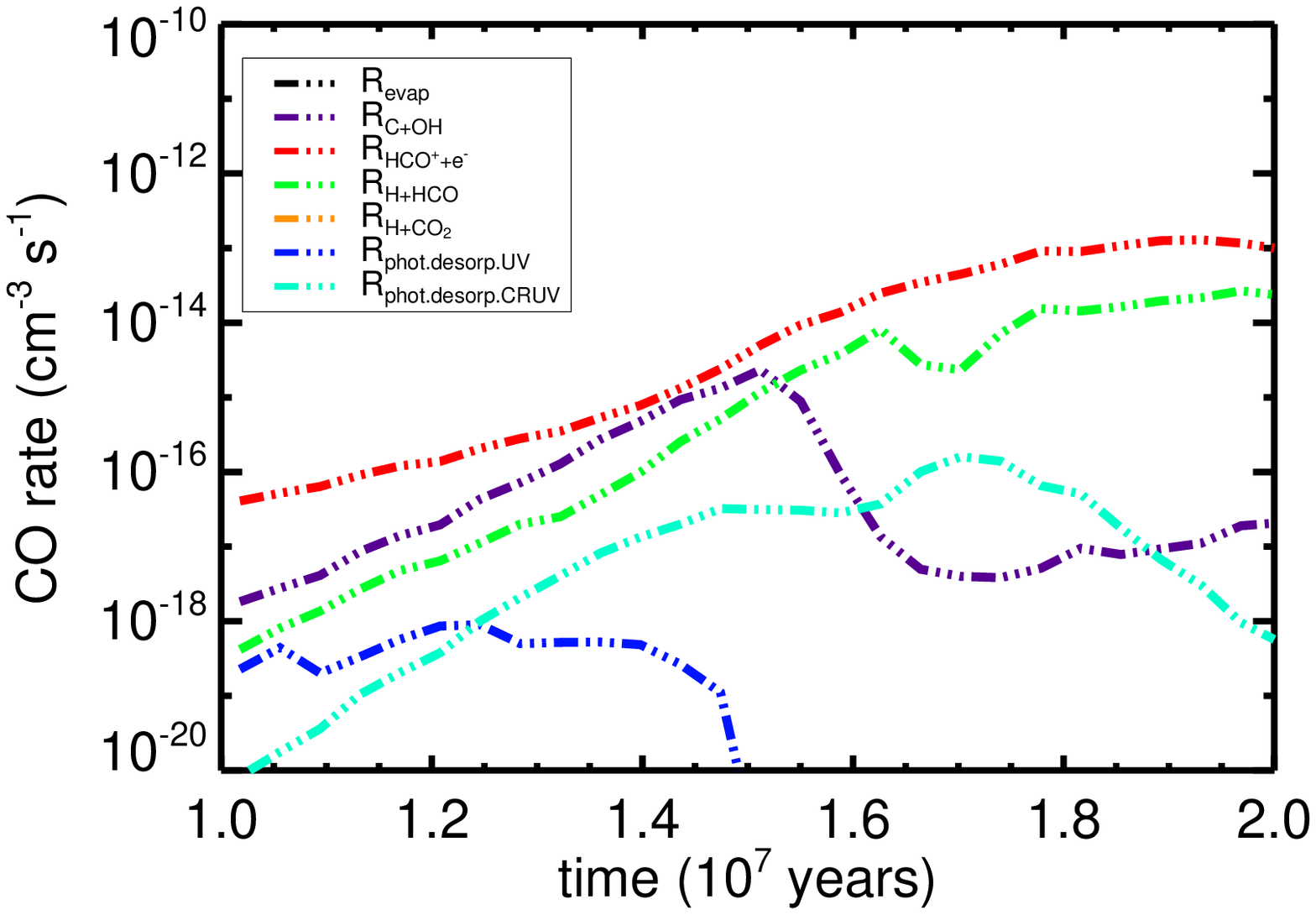} \\
\includegraphics[scale=0.51]{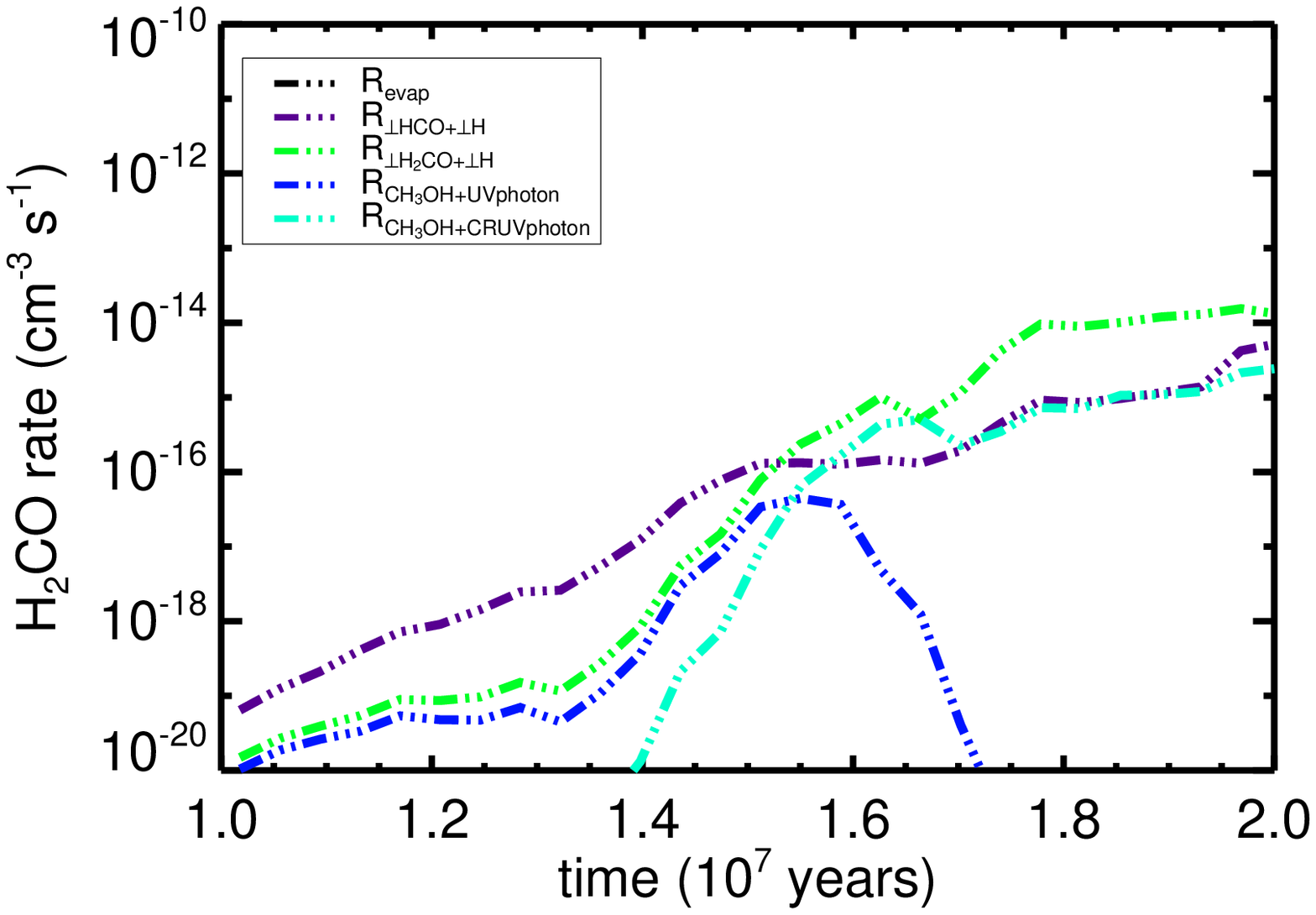}
\includegraphics[scale=0.51]{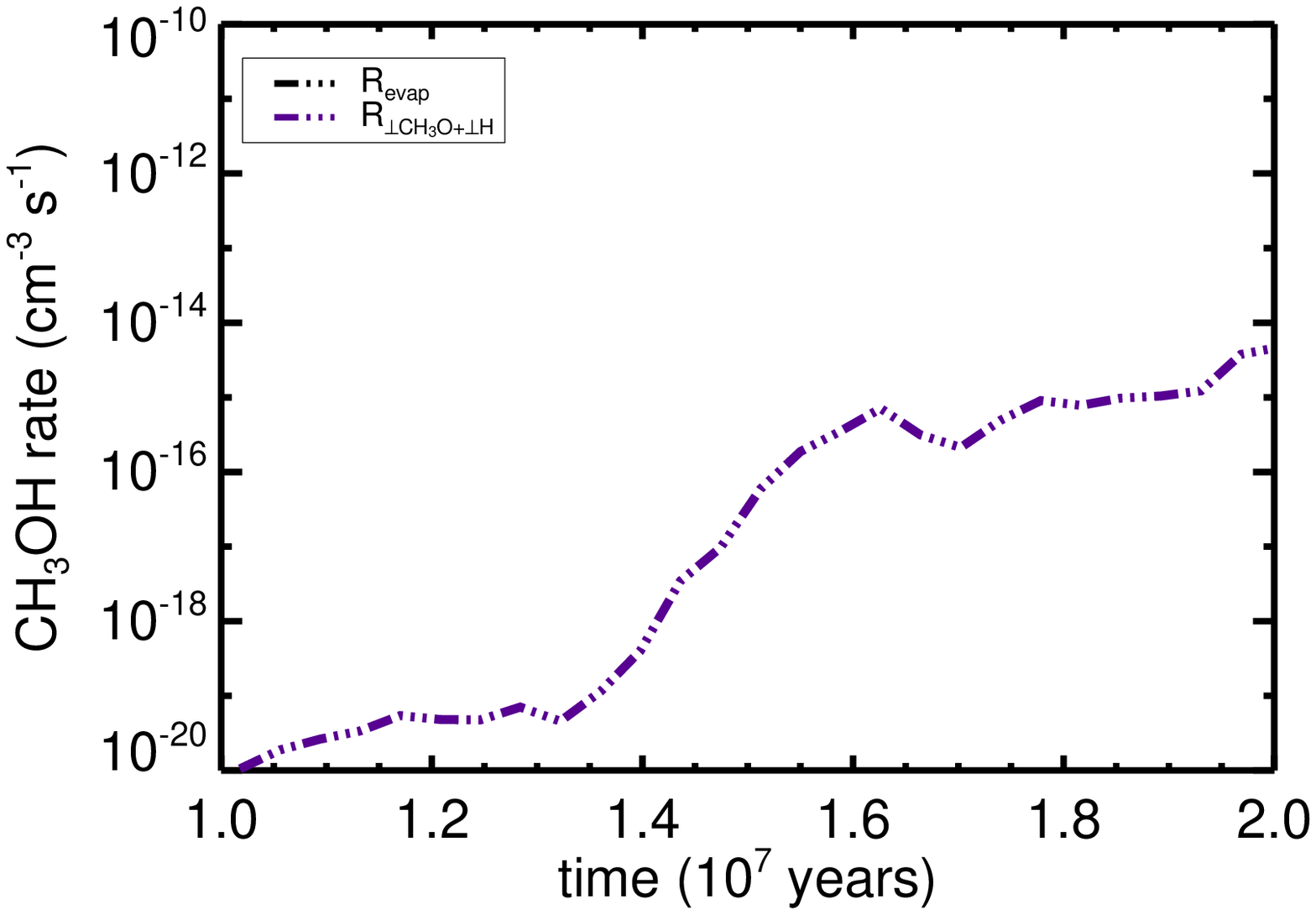}
\caption{Formation rates of gas-phase species. The main formation rates of H$_2$O (top left), CO (top right), H$_2$CO (bottom left), and CH$_3$OH (bottom right) are plotted as a function of time in $10^7$ yrs. R$_{\rm evap}$, the evaportation rate, is drawn for each specie, however, this rate is quite low for methanol such that the line does not fall within the limits of the plot. Other molecular reactions as well as photo-processes that result in the four mentioned species are given within the legend.}
\label{fig:ratesgas}
\end{figure*}

\begin{figure*}[htb!]
\includegraphics[scale=0.51]{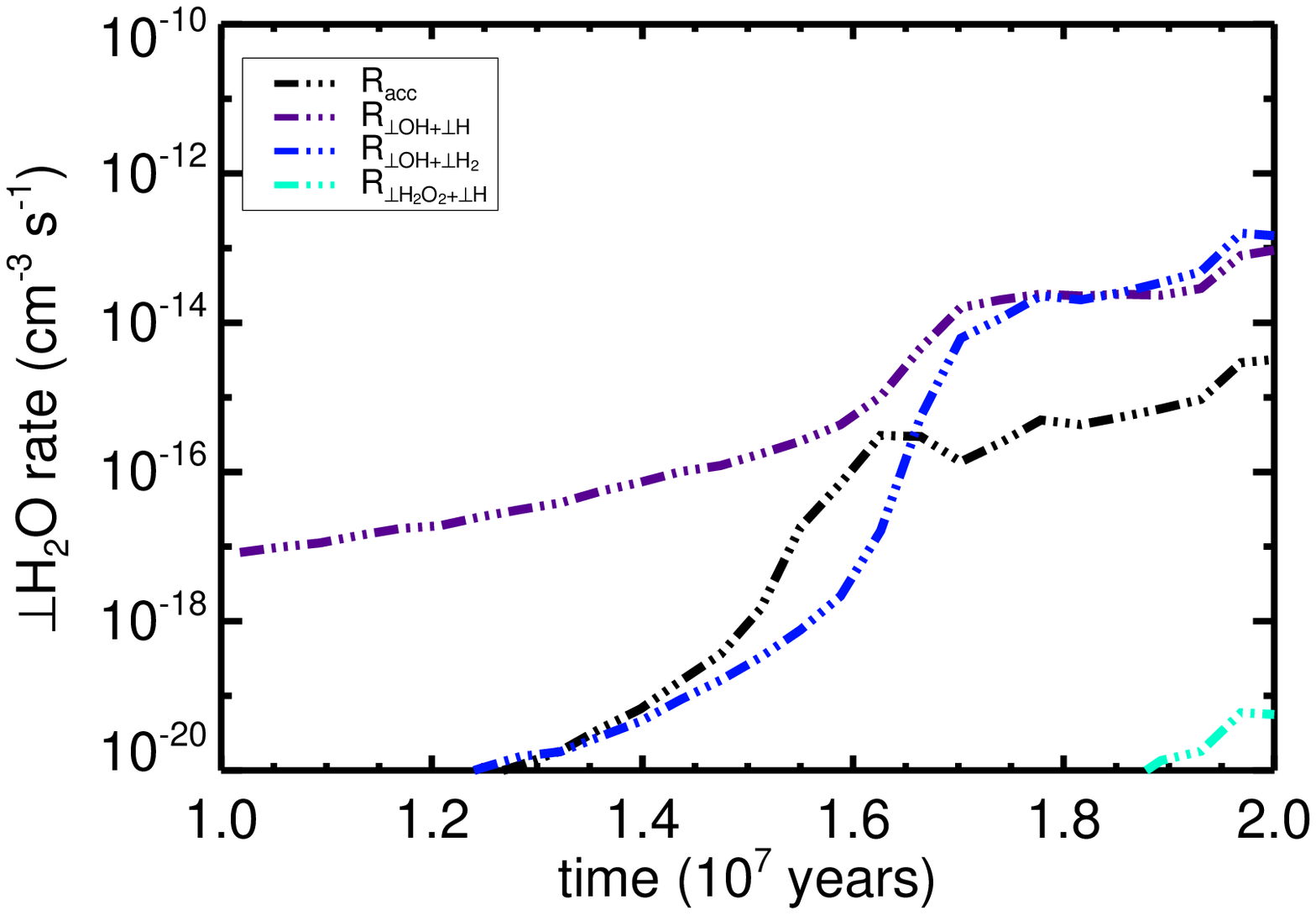}
\includegraphics[scale=0.51]{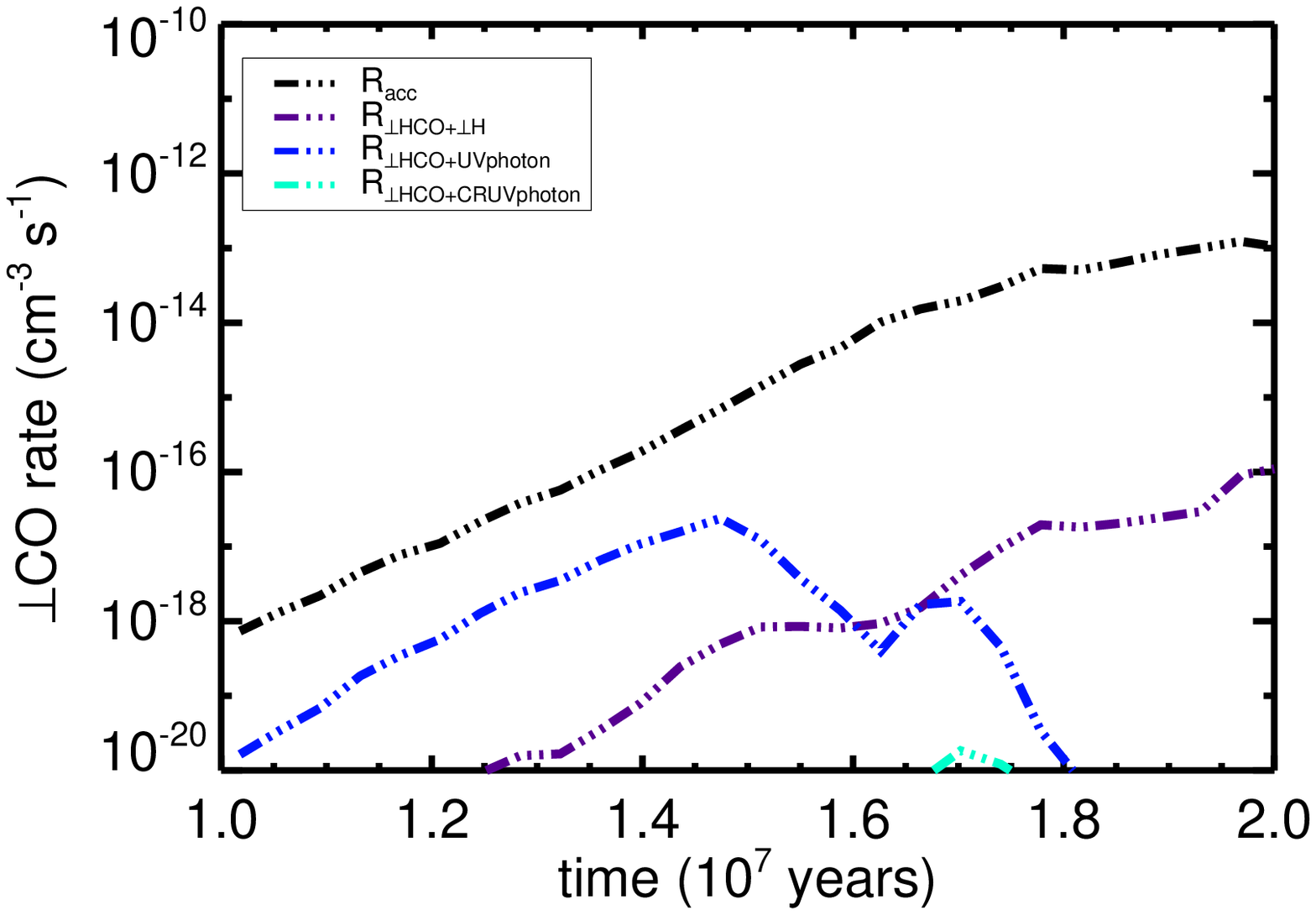} \\
\includegraphics[scale=0.51]{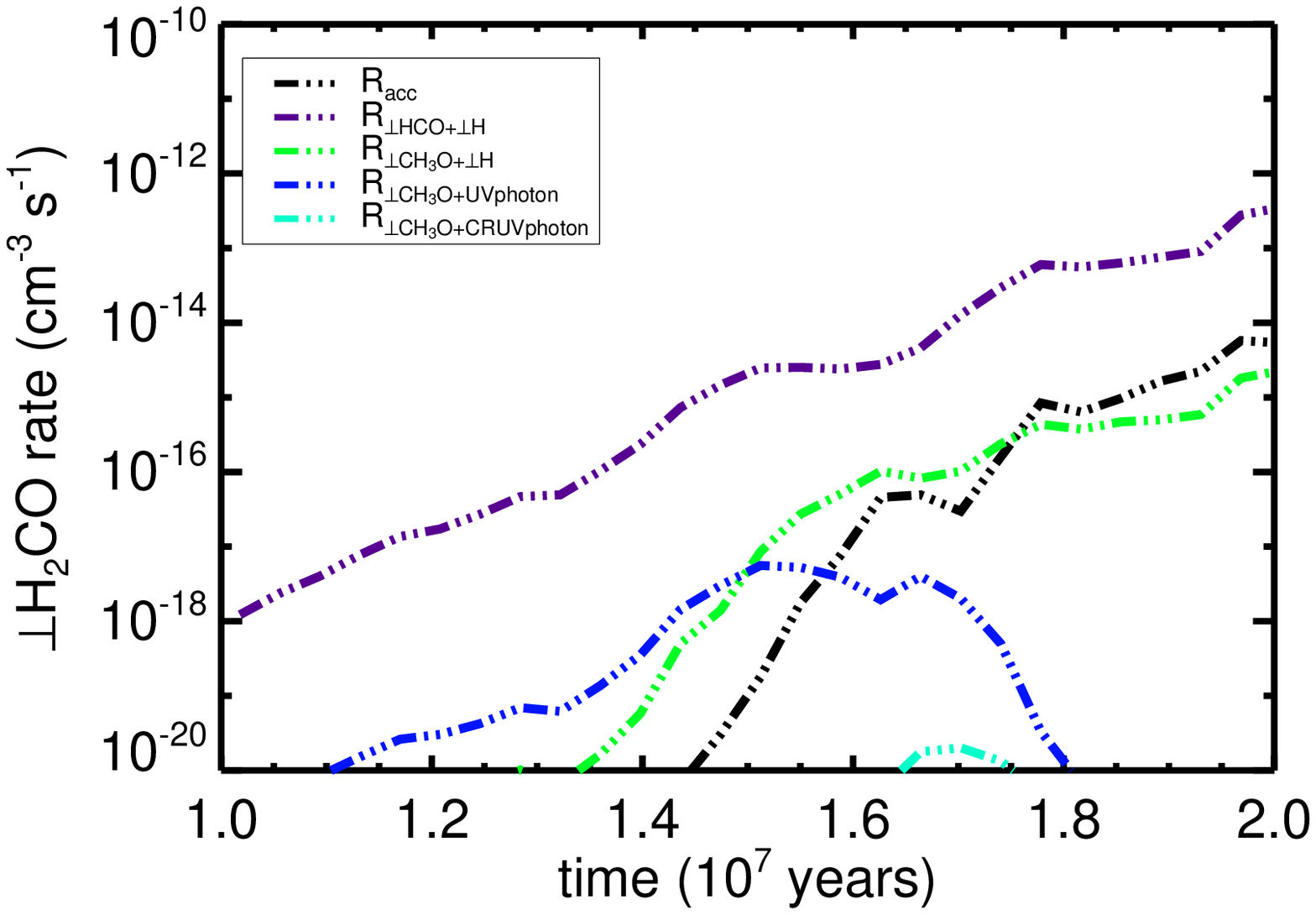}
\includegraphics[scale=0.51]{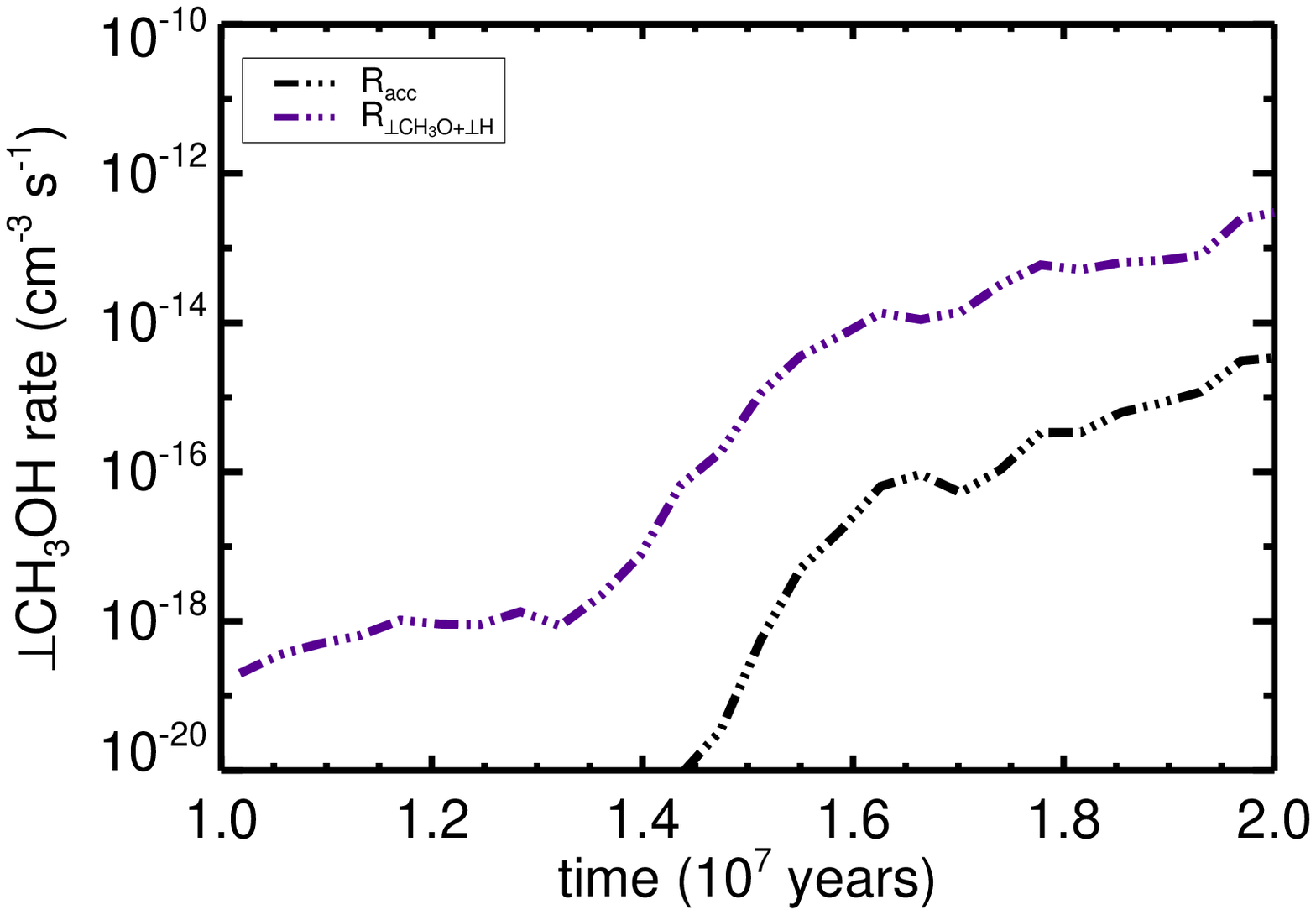}
\caption{Formation rates of ice species. The main formation rates of $\bot$H$_2$O (top left), $\bot$CO (top right), $\bot$H$_2$CO (bottom left), and $\bot$CH$_3$OH (bottom right) are plotted as a function of time in $10^7$ yr. Surface reactions and photo-processes that result in the four mentioned species are given within the legend.}
\label{fig:ratesice}
\end{figure*}

By following the formation rates of the species in our model, we can understand the preferred formation pathways for each of our species. Here we report the formation rates of the species H$_2$O, CO, H$_2$CO, and CH$_3$OH during collapse. In the following images, Figs. \ref{fig:ratesgas} (gas species) and \ref{fig:ratesice} (ice species), we plot the main formation rates for each of these species.

\noindent \\
\textit{Gas-phase formation rates}\\ %desorption 90, 15 --> 19, 18
In the top left panel of Fig. \ref{fig:ratesgas}, we see that gas-phase water has two main formation channels. One through the reaction of $\rm H_3O^+$ with $\rm e^-$ in the gas phase, the other through chemical desorption by the reaction of $\rm \bot OH$ with $\rm \bot H$ on dust surfaces from which 90\% of the reactions on bare grain surfaces desorb into the gas phase. This drops to 18\% when the surface is covered by ice. This means that if the dust formation channel is not considered, the gas-phase water formation rate will typically be underestimated by a factor of $\sim$2 with the conditions used in our simulations. Other channels, such as $\rm OH$ with $\rm H$, $\rm H_2$ with $\rm O^-$, and the second important desorption channel $\rm \bot H_2O_2$ with $\rm \bot H$ are less significant routes to form gas-phase water. The latter channel may become much more relevant if reaction-diffusion competition is considered for tunneling because of the high ($E_a = 1000$ K) barrier.

In the top right panel of Fig. \ref{fig:ratesgas}, we address the formation rates of CO in the gas phase. CO mainly forms through the dissociative recombination reaction $\rm HCO^{+} + e^{-}$. This recombination rate is quite high as long as there are enough electrons around. The CO formation rate is also strongly dependent on the HCO supply in the gas phase, in which, as we show below, surface chemistry plays a major role. Below \nhtot $= 10^3$ \cmcube, $\rm HCO^{+}$ is primarily formed by the reaction C$^+$ with H$_2$O, but at higher densities %, and in our simulations at $t > 1.5\times10^7$ yr, 
$\rm HCO^{+}$ is mainly produced through the ionization of chemically desorbed HCO. The second important formation route to form CO is through $\rm C + OH$, but only early on in cloud evolution, that is, when the cloud is still in a diffuse stage. The CO formation rate by $\rm C + OH$ sharply decreases after it peaks around t = 1.5$\times10^7$ yr. This is because atomic carbon is becoming depleted as more and more carbon is converted into CO and HCO. %The CO abundance at its peak is $\simeq 1.28\times10^{-4}$. 
Once atomic carbon is depleted, the remainder of O and OH can be channeled into other reactions. We can see, for example, an upturn in the water formation rate through the reaction of $\rm OH + H$ at the same moment in time. $\rm H + HCO$ becomes the second-most important channel at later stages ($\rm t > 1.55\times10^7$ yr) as we enter the translucent cloud stage (\nhtot $> 10^3$ \cmcube). Even though these are all gas-phase reactions, they are heavily affected by the production of HCO and OH on grain surfaces that are subsequently being released into the gas phase. Without the catalyzed species formation on dust, gas-phase CO formation rates will be constrained. We note that since hydrocarbon chemistry is beyond the scope of this work, an important CO formation channel in diffuse regions and at edges of molecular clouds, i.e., $\rm CH_2 + O$, is omitted \citep{1985ApJ...291..722T, 2008ApJ...683..238K}. Long carbon chain species, such as C$_9$H$^+$, which do not take part here, may also influence CO formation rates \citep{2002A&A...381L..13R}. 

After the formation of the first ice layer at t = 1.65$\times10^7$ yr, $\rm H$ and $\rm HCO$ are more strongly bound to the surface, see Table \ref{tab:bindingenergies} for adsorption energies. This reduces the mobility of atoms on the dust surface as well as the desorption probabilities because they both depend on the binding energy, which explains the acute momentary decline in the second-most important rate at this time, $\rm R_{H + HCO}$. We also see that photodesorption is not an important CO producer during the whole evolution of the cloud, from diffuse conditions to the first core formation. Even without our two-ice-layer penetration restriction for the UV photons, the photodesorption rates would still be an order of magnitude lower than the main CO formation rate. Since CO photodesorption occurs most efficiently by photons with energies of around 8-9 eV \citep{2010A&A...522A.108M, 2011ApJ...739L..36F}, while CO is photodissociated in the gas phase by UV photons with energies of $>$11 eV, we might be underestimating the photodesorption rate when taking into account CO self-shielding for photodesorption. A test run without any self-shielding showed that the photodesorption rate is then much higher, and almost equal to the rate $\rm R_{C + OH}$, but is still lower by more than one order of magnitude than the main CO formation rate $\rm R_{HCO^+ + e^-}$. This did not affect our results. Lastly, we can see that gaseous CO is not enhanced by thermal evaporation as the dust temperature is too cold $\rm T_{dust} = 7-8$ K for $\bot$CO to be released. In fact, this will eventually lead to the depletion of CO from the gas phase as accretion starts to become more important with increasing density.

%desorption 1.5, 5 --> 1, 5
%desorption 15, 50 --> 10, 50
In the bottom left panel of Fig. \ref{fig:ratesgas}, we display the formation rates of formaldehyde in the gas phase. The most dominant pathway to form formaldehyde in diffuse clouds ($\rm t < 1.5\times10^7$ yr, \nhtot $< 10^3$ \cmcube) is by direct chemical desorption of $\bot$H$_2$CO following the reaction $\bot$HCO + $\bot$H. Only 1\% (ice) to 5\% (bare) of this reaction results in the formation of H$_2$CO (gas), but it still emerges as the dominant rate. \cite{2013A&A...560A..73G} also concluded that the gas-phase formaldehyde formation is due to nonthermal desorption, but attributed its production to photodesorption. \cite{2009A&A...498..161V} affirmed the need of a continuous supply of formaldehyde, for instance, from grain surface chemistry, to explain the abundances in their observations of the Orion Bar. Above t = 1.55$\times10^7$ yr, chemical desorption following the reaction $\bot$H$_2$CO + $\bot$H, together with the photodissociation of methanol in the gas by CRUV photons, overtakes the former ($\rm R_{\bot HCO + \bot H}$) desorption rate. In 10\% (ice) to 50\% (bare) of the cases, the reaction $\bot$H$_2$CO with $\bot$H results in the desorption of the products. This reaction leads to the formation of a $\rm CH_3O$ radical that may be released into the gas only to quickly form H$_2$CO + H. The in-between steps are omitted here, since they are relatively fast. Thermal desorption of $\bot$H$_2$CO is not a significant gas-phase supplier for formaldehyde either.

In the bottom right panel of Fig. \ref{fig:ratesgas}, we show that methanol has only one essential formation mechanism. This is by chemical desorption from the grain surface reaction of $\bot$CH$_3$O + $\bot$H. The rate remains below $10^{-14}$ \cmcube s$^{-1}$, which means that it will not result in much methanol into the gas phase in a cloud lifetime of $\sim 10^{14}$ s. Most of the methanol will remain frozen onto grain surfaces. The gas can only be enriched by methanol through evaporation of the frozen species at higher temperatures inside the cloud.

\noindent \\
\textit{Ice species formation rates}\\ %desorption 10, 85 --> 10, 82
In the top left panel of Fig. \ref{fig:ratesice}, we present the formation rates of water on surfaces. Water ice primarily forms by the reaction $\rm \bot OH + \bot H$. This exothermic reaction occurs without a barrier and has a high chance to chemically desorb. The percentage that remains on the dust surface is 10\%, which increases to 82\% for an icy substrate. This effect is directly visible in the figure from the jump in reaction rates at t = 1.65$\times10^7$ yr when the dust is covered by an ice mantle. The reaction $\rm \bot OH + \bot H_2$ also increases greatly upon first ice layer formation. With 100\% of the products remaining on dust, this reaction has a barrier of $E_a = 2100$\,K that needs to be overcome. Despite this, it dominates the water-ice formation rate at later times when the cloud density is higher, that is, \nhtot $\geq 10^{4}$ \cmcube at the dense molecular stage. This is mainly due to the increasing $n_{\rm H_2}/n_{\rm H I}$ ratio during cloud evolution. Accretion of gaseous water becomes important at densities of \nhtot $\simeq 4\times10^{3}$ \cmcube, with gas and dust temperatures of around 10 K. By forming $\rm H_2O$ on dust, releasing it into the gas and reaccreting it back onto the dust will make the ice more porous. %REFERENCE?
After fully covering the dust surface with water ice, the accretion rate decreases like the gas-phase water formation did as a result of the decline in the desorption probability of the main formation rate.

In the top right panel of Fig. \ref{fig:ratesice}, we give the formation rates of $\bot$CO. CO ice grows on grain surfaces predominantly by accretion. The high CO abundance in the gas phase and the low gas and dust temperatures that allow for a high sticking coefficient result in a high accretion rate. The CO ice formation mechanism becomes somewhat circular above t = 15 Myr, because CO on grain surfaces are hydrogenated to form HCO, thereupon to be chemically desorbed into the gas phase. The desorbed HCO molecule is quickly ionized through photoionization or ion exchange reactions to form HCO$^+$, making it the primary route to form HCO$^+$. This ion, as we know, dissociates into CO in the gas phase only to be reaccreted on dust grains where the whole cycle restarts. A demonstration of the cycle is given below.
\begin{displaymath}
\rm CO_{gas} \stackrel{a}{\rightarrow} CO_{ice} \stackrel{b}{\rightarrow} HCO_{ice} \stackrel{c}{\rightarrow} HCO_{gas} \stackrel{d}{\rightarrow} HCO^{+}_{gas} \stackrel{e}{\rightarrow} CO _{gas}
\end{displaymath}
a = accretion \\
b = hydrogenation \\
c = chemical desorption \\
d = ionization \\
e = dissociative recombination \\
%
% \begin{eqnarray}
% \rm CO_{ice} &\rightarrow& \rm HCO_{gas} \rightarrow HCO^{+}_{gas} \rightarrow CO_{gas} \rightarrow CO_{ice} 
% \\
% \rm accrete &\rightarrow& \rm desorb \rightarrow ionize \rightarrow dissociate \rightarrow accrete
% \nonumber
% \label{eq:cocycle}
% \end{eqnarray}
%

The chicken-and-egg problem is circumvented since CO initially forms in the gas phase through a channel independent of itself, namely $\rm H_2O + C^+ \rightarrow HCO^+ + H$ at \nhtot $< 10^3$ \cmcube. 
%HCO$^+$ in the gas phase in diffuse and translucent clouds (\nhtot $< 10^3$ \cmcube) is predominantly formed through the reaction H$_2$O + C$^+$. 
This route is still associated with surface chemistry to a certain degree because gas-phase water formation is enhanced by the reactions on dust grains. A relatively high CO abundance is sustained in the gas phase by the continued supply through the cycle at low, $T_d < 10$ K, temperatures, which would otherwise result in the rapid freeze-out of CO. Photodissociation by UV photons at low extinction $A_V < 3$, and the dissociative reaction $\rm \bot HCO + \bot H$ at high extinction $A_V > 5$ are other, but minor producers of $\rm \bot CO$. Since the CO freeze-out occurs throughout cloud evolution, it is expected that CO ice will be present and well mixed within every ice layer covering the grain surfaces. The CO in the upper layers will subsequently become more and more hydrogenated as the medium density rises. This will decrease the amount of CO ice, but the first ice layer(s) should still have pristine CO ice mixed with water ice.

%desorption 95, 98.5 --> 95, 99
In the bottom left panel of Fig. \ref{fig:ratesice}, we report the rates needed to form $\bot$H$_2$CO. Formaldehyde ice mainly forms after two successive hydrogenations of $\rm \bot CO$. The reaction $\rm \bot HCO + \bot H$ is, therefore, the main producer of formaldehyde ice throughout cloud evolution. This reaction has a 95\% (bare) to 99\% (ice) probability to let the product remain on the surface of the dust grains. Reaccretion of formaldehyde by the chemically desorbed $\rm \bot H_2CO$ from the this reaction, as this is one of the main gas-phase suppliers, is low because of the low desorption percentages. De-hydrogenation (removing an H atom) of $\rm \bot CH_3O$ also produces by about two orders of magnitude less formaldehyde ice than the main rate. This is because the reaction $\rm \bot CH_3O + \bot H \rightarrow \bot H_2CO + H_2$ has a reaction barrier, albeit low, of $E_a = 150$ K that needs to be overcome.

%desorption 95, 98.5 --> 95, 99
The bottom right panel of Fig. \ref{fig:ratesice} shows the formation rates of CH$_3$OH on surfaces. Methanol ice essentially forms by four repeated hydrogenations of CO. In a similar fashion as formaldehyde, methanol ice forms through the exothermic reaction $\rm \bot CH_3O + \bot H$ with a 95\% (bare) to 99\% (ice) probability to make $\rm \bot CH_3OH$. The formation rate is initially low because first $\rm \bot CH_3O$ has to be created in ample amounts. The rate increases after t = 1.33$\times10^7$ yr when $\rm \bot CH_3O$ is steadily formed, and this rate is similar to that of formaldehyde. Methanol will still be more abundant than formaldehyde since it is a more stable, strongly bound molecule. To break up methanol with a neutral hydrogen atom, a barrier of $E_a = 3000$ K needs to be overcome, which allows it to survive and build up over time. Methanol is also being reaccreted by the molecules that were initially created on dust and were immediately desorbed into the gas. 

\begin{figure}[htb!]
\includegraphics[scale=0.51]{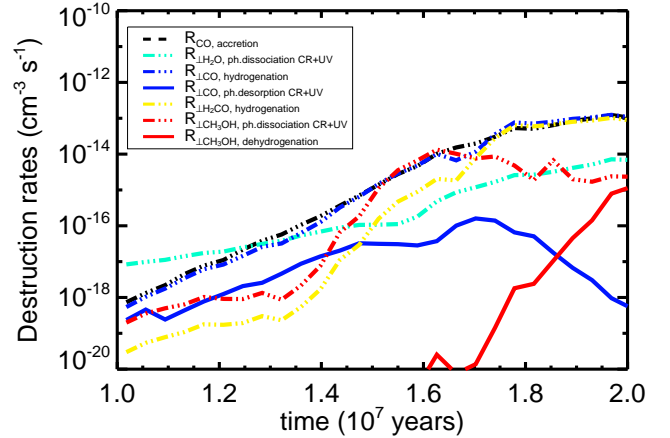}
\caption{Main destruction rates of key species. We plot the important destruction rates of H$_2$O (ice), CO (gas+ice), H$_2$CO (ice), and CH$_3$OH (ice) as a funtion of time. The destruction of CO through surface accretion is drawn in black, $\bot$CO through hydrogenation and photodesorption in dark blue, $\bot$H$_2$O through photodissociation in light blue, $\bot$H$_2$CO through hydrogenation in yellow, and $\bot$CH$_3$OH through photodissociation and dehydrogenation in red. For photo-processes, the rates obtained from cosmic rays and UV photons are added up.}
\label{fig:destructionrates}
\end{figure}

A caveat in our chemical network is that we did not include species larger than methanol, nor did we treat all the associated ions of methanol in the gas phase. Even though the gas-phase chemistry is not expected to be as important as grain surface chemistry, this will still create a sink out of methanol and therefore make us overestimate it, especially at later times.

\noindent \\
\textit{Destruction rates}\\ 
The most important destruction rates for the species discussed in this section are displayed in Fig. \ref{fig:destructionrates}.
For the gas phase, only the accretion rate of CO is presented in this figure to highlight the competition between the destruction rate of $\bot$CO. Other accretion rates are shown in Fig. \ref{fig:ratesice}. For ice species, the dominant destruction rates are given with the addition of CO photodesorption and methanol dehydrogenation.

The CO that is accreted is quickly hydrogenated to form HCO. At t = 16.5 Myr the hydrogenation rate diminishes somewhat, which will result in the increase of $\bot$CO. The accretion rate of CO also slows down shortly thereafter because the chemical desorption of HCO that influences the CO abundance in the gas phase is also reduced. In the end, a new balance is reached. Like CO, formaldehyde ice is mainly destroyed by hydrogenation. The rate is initially lower than that of CO due to the lower abundance of $\rm \bot H_2CO$. That the two hydrogenation rates are equivalent at later times suggests that a balance is reached between formation and destruction. Methanol ice is mostly destroyed by radiation. Only at very late times, dehydrogenation of methanol becomes the strongest destruction mechanism. This is due to the high activation barrier ($E_a=3000$\,K) in the dehydrogenation of methanol.

The main point is that all rates exhibit a twist in their curves at the time when ice completely covers the dust surface (t = 16.5 Myr). Following the transition in surface binding energies of species at this time, the reaction rates on grain surfaces change, that is, the mobility is reduced when the binding energies increase, while the chemical desorption probabilities decrease when the surface is covered by water ice.

\subsection{Ice distribution around the clump}
\begin{figure*}[htb!]
\includegraphics[scale=0.51]{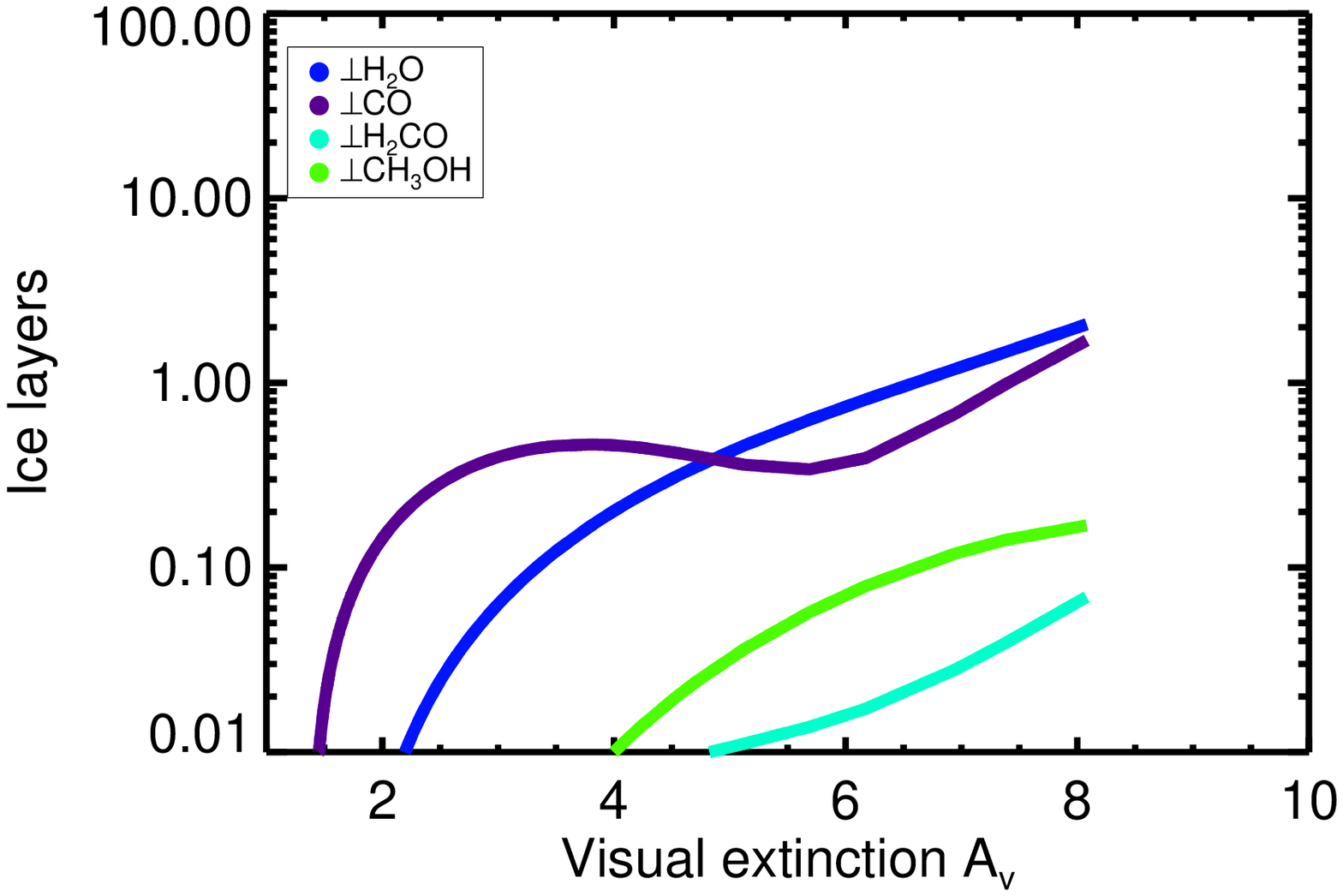} %\\
\includegraphics[scale=0.51]{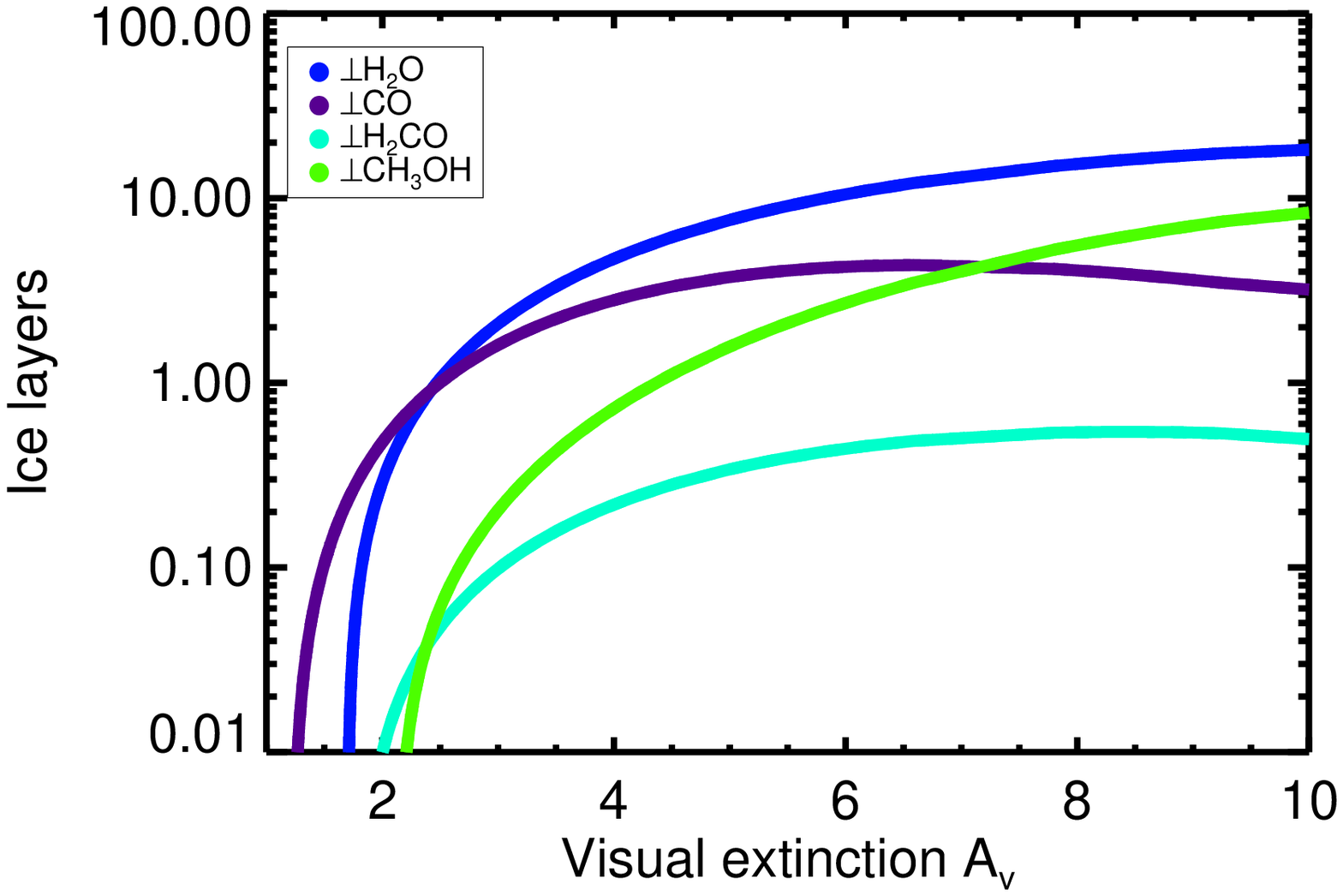}
\caption{Distribution of ices in and around a molecular clump. The amount of ice layers covering the surface of dust for several species is plotted as a function of visual extinction. The left panel displays the environment of a clump at a time of t = 1.66$\times 10^7$ yr of cloud evolution. The right panel shows a time snapshot of the clump after t = 1.82$\times 10^7$ yr of cloud evolution. The colored curves are least-squares fits to simulation data points.}
\label{fig:rcore1}
\end{figure*}

We also examined the abundance profiles of the ices within a radius of 5 pc from the clump center. Inside this region, we followed the growth of ices on dust surfaces as a function of optical depth and present column density maps. We inspected the optical depth behavior for two different epochs, at a time of t = 16.6 Myr and t = 18.2 Myr after we started our simulation. In the period between these two time intervals, the transition from CO ice to methanol ice occurs inside the clump. The density of the clump center for the two snapshots is \nhtot = $4\times10^3$ \cmcube and $10^4$ \cmcube, respectively, while the gas temperature in both time frames lingers around 11 K (see Fig. \ref{fig:clumpvar}). The gas temperature in the border region ($A_V \lesssim 1$) is higher T$_g$ $\geq 20$ K, while the dust temperature is low T$_d$ $\sim 7$ K. Our resolution limits us in mapping the inner pc of our clump, and we did not have many data points at the center. This makes our curves appear rather smooth. A least-squares fifth-order polynomial was fitted to our data points and with this, the ice distributions for the two epochs are displayed in Fig. \ref{fig:rcore1}.

A thicker layer of ice covers the surface of the dust in the inner parts of the clump where $A_V$ and density are higher. Solid CO is more extended than solid water at a surface coverage below one mly. $\bot$CO, however, also strongly contributes at the center of the clump. This establishes that water and CO ices are well mixed everywhere in the clump during the first ice layer formation. Clearly water ice becomes the main ice constituent of the icy mantle when one mly of water ice has formed. The transition at which solid water becomes more important than solid CO occurs at $A_V=4.8$ during the translucent cloud stage (see left panel of Fig. \ref{fig:rcore1}), while 1.6 Myr later, at a dense molecular stage, this transition takes place at $A_V=2.3$ (see right panel of Fig. \ref{fig:rcore1}).
%At earlier epochs, this occurs at a higher $A_V$ than in later epochs. E.g., in the left panel of Fig. \ref{fig:rcore1}, we see this transition occurring at $A_V=4.8$, while in the right panel, i.e., 1.6 Myr later, this transition takes place around $A_V=2.3$. 
Deeper inside the core, methanol and formaldehyde ice is rapidly being formed at the loss of $\bot$CO. The methanol ice surface coverage surpasses the CO ice coverage around an $A_V$ of 7 at t = 1.82$\times 10^7$ yr, but this transition also shifts to lower $A_V$ at later times. For example, the transition occurs at $A_V=4.5$ at t = 2$\times 10^7$ yr. 

From the difference between two time frames, we can understand that the composition of the ice mantles do not strongly depend on the optical depth, since we have different compositions at different times for a given value of $A_V$. We can infer from this that photo-processes with a strong dependence on optical depth, such as photodissociation and photodesorption, are not significant factors to the ice composition within the conditions used in this work.

\begin{figure*}[htb!]
\includegraphics[scale=0.32]{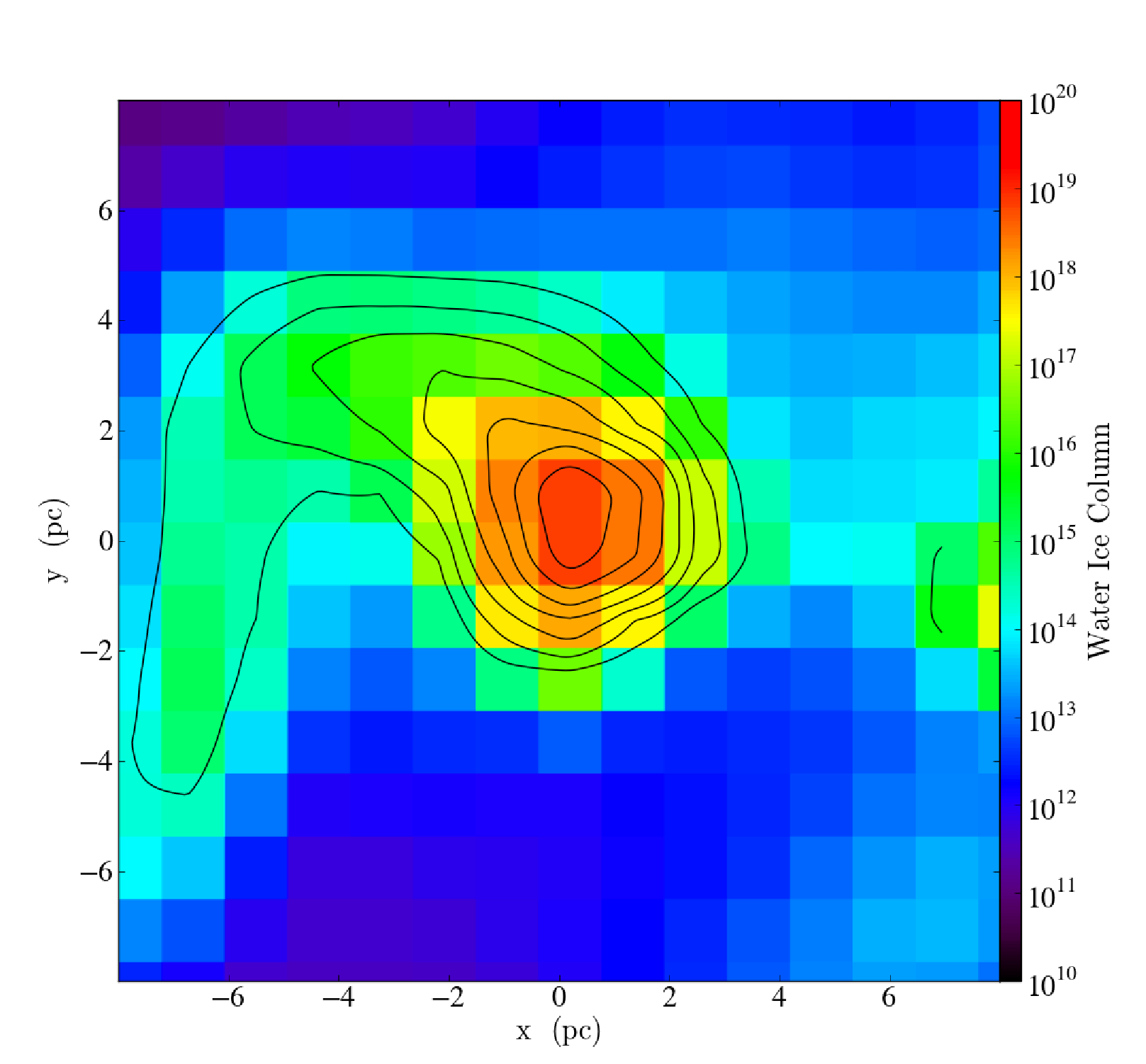} %ARXIV
\includegraphics[scale=0.32]{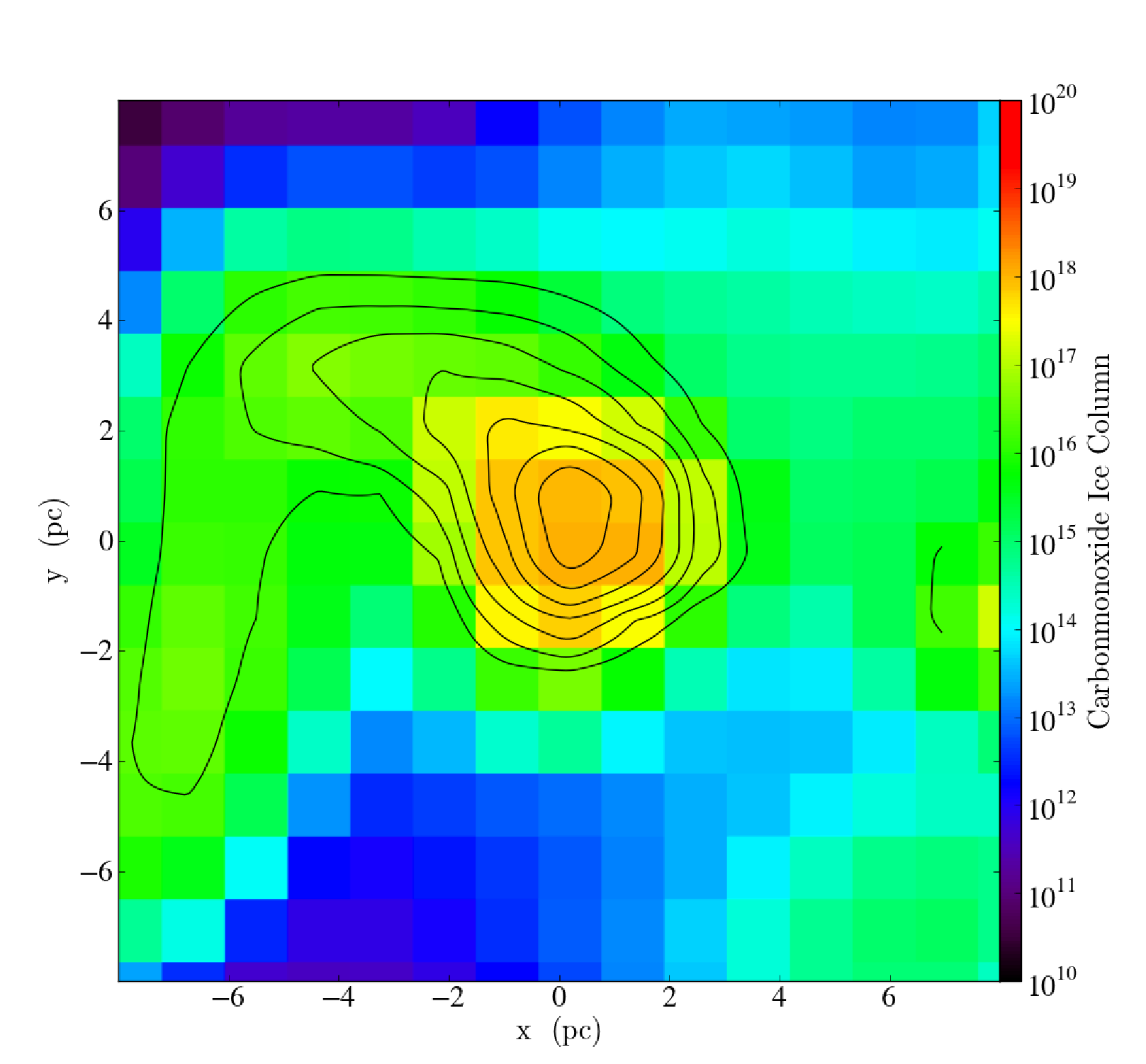} \\
\includegraphics[scale=0.32]{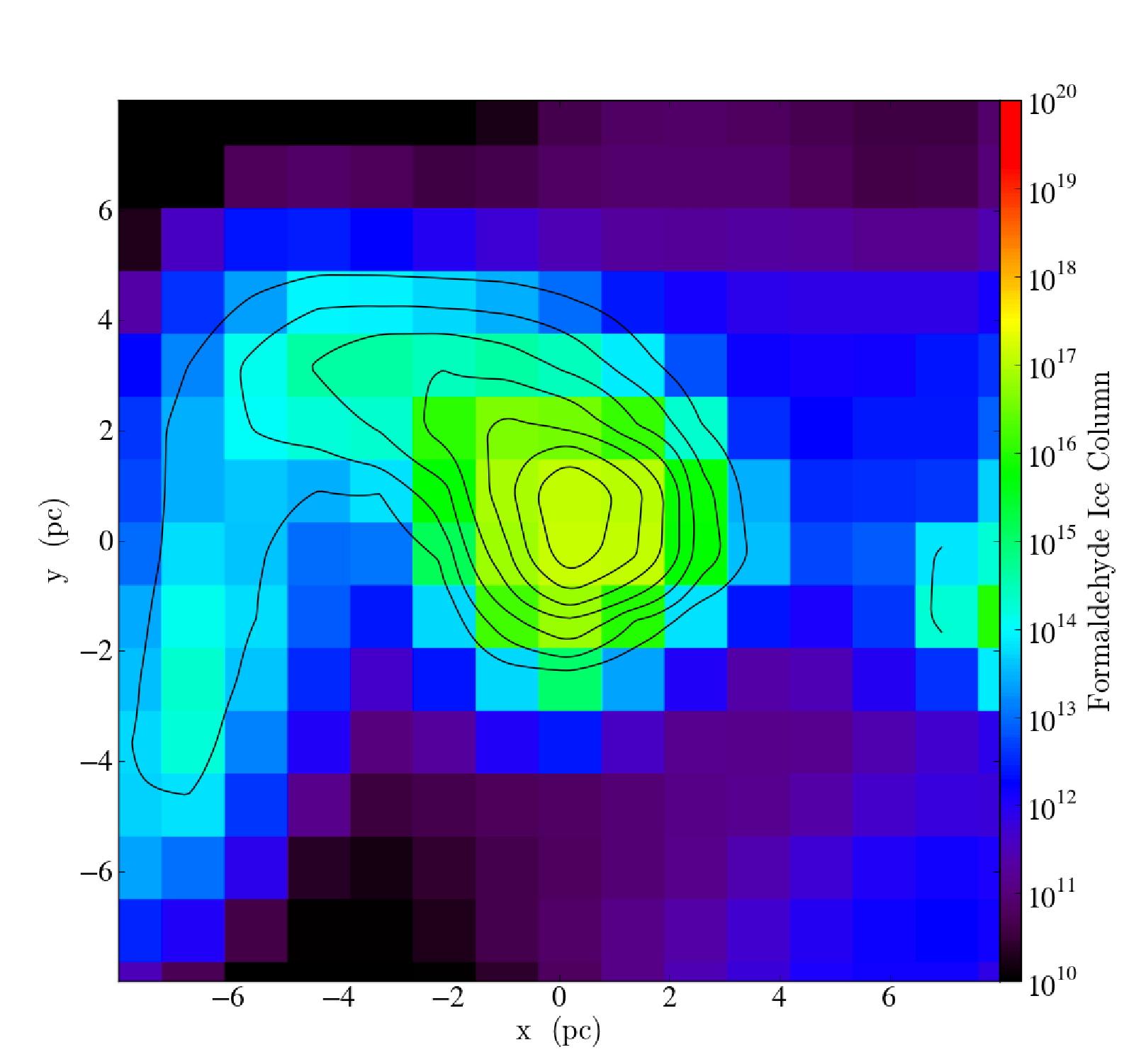}
\includegraphics[scale=0.32]{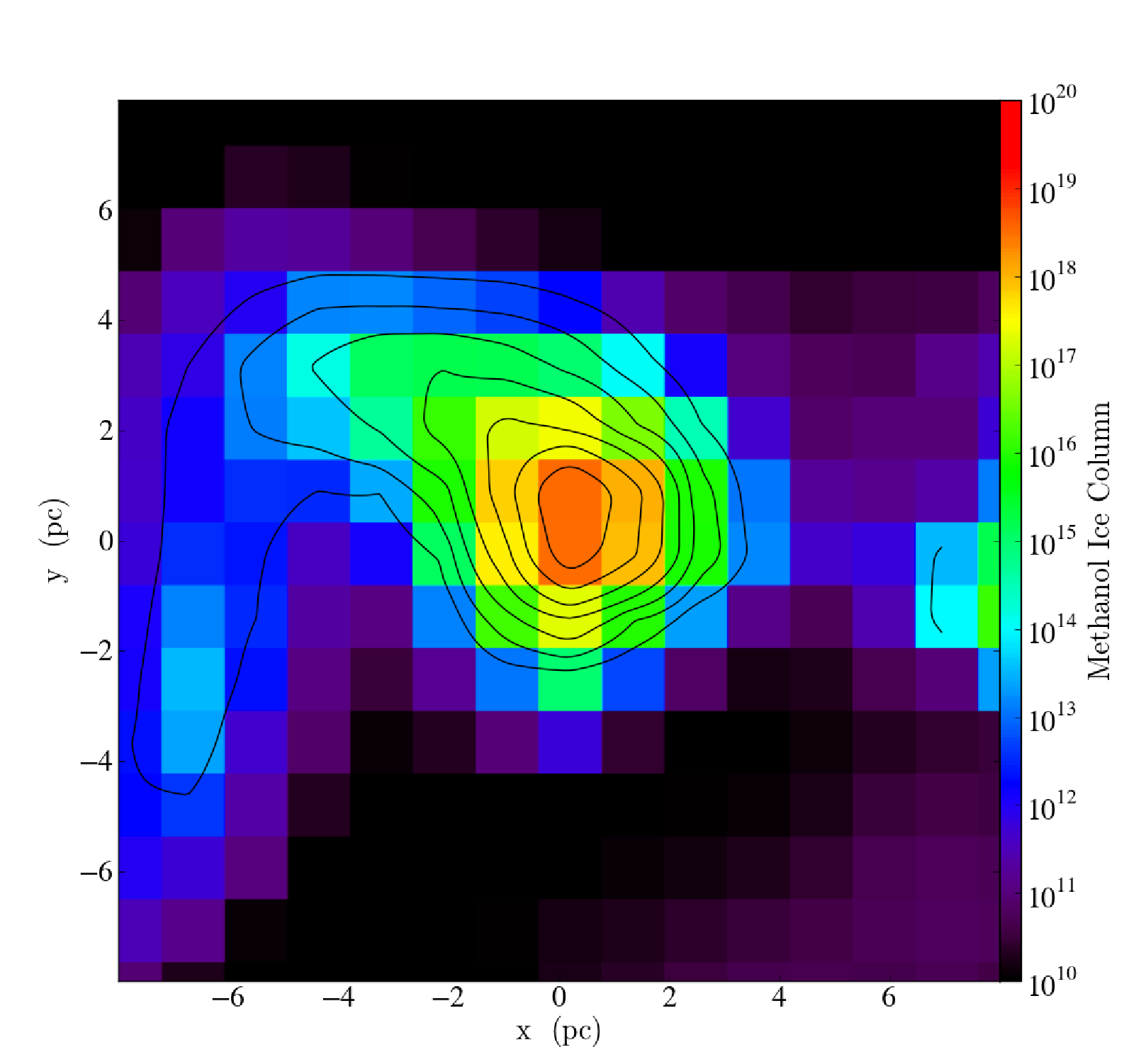}
\caption{Ice maps. In these images, a projection along the $Z$-axis, resulting in a line-of-sight column (cm$^{-2}$), is plotted within a 16x16 pc box. From top left to bottom right, ice maps of the species $\bot$H$_2$O, $\bot$CO, $\bot$H$_2$CO, and $\bot$CH$_3$OH are displayed. The color range representing the species column densities spans from low 10$^{10}$ (black) cm$^{-2}$ to high 10$^{20}$ cm$^{-2}$ (red) values. The contours enclose the region where the total column density, $N_{\rm H} = \int n_{\rm H} ds$, ranges from $10^{22}$ to $6\times10^{22}$ cm$^{-2}$, i.e., $A_V = 4.5 - 27$, to indicate where most of the matter lies.}
\label{fig:maps}
\end{figure*}

We also present the distribution of ices in spatial dimensions. In Fig. \ref{fig:maps} we show maps of the species as they would appear on the sky. 
Here we plot the line-of-sight column densities of the species in and around the dense clump. These maps were created at the time of t = 1.82$ \times 10^7$ yr after simulation start, which is the same time as in the right panel of Fig. \ref{fig:rcore1}. The column densities range from 10$^{10}$ cm$^{-2}$ to 10$^{20}$ cm$^{-2}$ for all species. Water has the strongest peak at the center, followed by methanol. $\bot$CO has a weaker central feature, but is more extended. Most of the cells are colored green with columns of around $N_{\rm CO} = 10^{16}$ cm$^{-2}$. This tells us that CO ices will be present for a wide range of environmental conditions. Methanol, on the other hand, is very centralized with a high peak at the center, but with very little methanol ice extending outward. At this stage, water ice and methanol ice emission lines should be the easiest to detect; they have column densities of 10$^{19}$ cm$^{-2}$ and above inside the clump.

% Methanol is the most stable of the series and does not react form higher-order complex molecules at typical ISM densities and temperatures. Destruction of methanol is mainly realized through photon processes. This results in a balance between $\bot$CO ice and methanol ice rather than converting everything into methanol. Deeper into the cloud, at higher opacities, and further into the evolution of the clump, photodissociation by UV photons decreses significantly thereby allowing the methanol reactions to become nearly one sided toward the formation of methanol ice. Only CRUV photons can still destroy methanol ice at low temperatures. At t = 1.1 t$_{\rm ff}$ ?, when the density of the cloud is \nhtot ? $\times 10^3$ \cmcube and the temperature is ?? K, we start seeing more methanol ice than CO ice. Methanol ice grows to a maximum of 24? monolayers until its growth is restrained. This amounts to 43\% ? of the total ice coverage. Water ice has nearly 56\% ? coverage.
% 
% %About observation of methanol and how much they find. We find too low CO2, but higher energies may be needed.
% 

\section{Summary of conclusions and discussion}
\label{sec:conclusion}
We performed for the first time hydrodynamical simulations of a collapsing gas cloud with detailed gas and grain surface chemistry in which the interplay between gas and dust is interlinked with the thermodynamics of the cloud. We presented our results on the impact of dust chemistry on the formation of ices and the enrichment of gas-phase species during the evolution of a gas cloud. We also revealed the dominant formation routes for several species at different stages of cloud evolution.

To obtain our results, we followed a gas cloud from a fully diffuse atomic stage (\nhtot $= 10$ \cmcube), which contracted and underwent phase transitions to finally form molecular clumps (\nhtot $> 10^4$ \cmcube). During the simulation, thick, $n_{mly} = 59$, ice layers formed within the densest clump. We find that the first ice layer covering the surface of the dust has a strong presence of adsorbed CO. This mainly comes from accretion of CO from the gas phase, since gaseous CO is in ample supply. However, to form CO in the gas phase, the gas cloud needs to be supplied with H$_2$O, HCO, and OH in the first place (CH$_2$ not included in this work). During the translucent and molecular cloud stages, these species are mostly formed on dust grains to be subsequently released into the gas phase through the process of chemical desorption. Without the enhanced HCO formation in translucent clouds through the exothermic reaction $\bot$H + $\bot$CO at \nhtot $\geq 10^3$ \cmcube, CO in the gas phase would form at a much lower rate. We conclude from this that gas-phase CO formation requires grain surface reactions to be effective when starting from fully atomic conditions. We also conclude that in this chemical desorption is essential in supplying HCO to the gas, but note that the desorption rate is sensitive to the activation barrier of the reaction.%, and the enhanced OH formation through $\bot$H + $\bot$O

Our results show us that the first ice layer is formed during the translucent cloud stage at a density of \nhtot $= 4\times 10^3$ \cmcube. CO ice is well mixed at this stage with water ice. After one mly of ice has formed, freeze-out occurs more rapidly as a result of the change in binding energy of the species with respect to an icy surface. This influences the mobility of species on surfaces as well as the chemical desorption probabilities, which eventually causes the formation of more water ice. We also see that CO ice gradually decreases with increasing density and opacity because it increasingly is more hydrogenated through the successive reactions with $\bot$H to form formaldehyde and methanol.

From the distribution of ices in the region surrounding the clump we see that most of the ices are at the clump center. CO ice is more extended toward the outer regions than water ice, and it is consumed in the clump center to form formaldehyde and methanol. Methanol favors the high-density regions, where it becomes the second-most abundant ice after water ice.

We conclude that grain surface chemistry strongly affects the abundances of species in the gas phase by creating strong, sometimes dominant, pathways for forming key molecules, such as water, formaldehyde, and methanol. Gas-phase species are also depleted by freeze-out onto dust surfaces at different cloud evolutionary stages. Water and especially CO freeze-out is seen to occur, which is confirmed by rates. Dust chemistry also allowed us to follow the formation and build-up of ices during the evolution of a gas cloud. Water ice becomes the dominant constituent of the ice mantle after 15.5 Myr of cloud evolution when the density has risen above $2\times10^3$ \cmcube, and methanol ice has a strong $\sim 43$\% presence inside dense clumps. These chemical influences affect the thermodynamic properties of the progenitors of star-forming regions. Moreover, we note that these changes occur and have an impact at early, translucent cloud stages that evolve into molecular clouds. These are critical moments for cloud fragmentation. \\

\noindent
In summary, our results are that
\begin{itemize} \itemsep8pt %4pt or 2pt
\item in the first ice layer(s), CO ice is well mixed with H$_2$O ice;
\item freeze-out of species greatly increases after one mly of ice covers the dust surface,
\item chemical desorption from grain surfaces can be important in supplying the gas phase with CO and formaldehyde;
\item methanol and formaldehyde (gas) is seen in higher density regions (\nhtot $>$ 4$\times 10^3$ \cmcube), while formaldehyde ice and CO ice are more likely to be found in the surrounding area; and finally, 
\item surface chemistry alters the species abundances in translucent clouds, which will affect the whole cloud evolution.
\end{itemize}

\subsection{Discussion}
Our chemical network consists of 42 species and 257 reactions. Although this is a respectable amount given the complexity of this work, the network is still restricted. In this, methanol is the largest molecule that we allow to form in our network. This will inevitably create a sink out of this molecule and makes us overestimate the methanol abundance since there are no routes to form higher order molecules or ions. %However, larger molecules are not expected to be present in copious amounts in such environments.

Chemical desorption following the hydrogenation of CO and formaldehyde is sensitive to the activation barriers we used. The activation barriers for the reactions $\rm CO+H$ and $\rm H_2CO+H$ are sometimes found to be high, $\rm 2000-2500\,K$ \citep[e.g.,][]{2008ApJ...682..283G, 2013JChPh.139p4310P}. Such high activation barriers would cause much less CO hydrogenation and chemical desorption to occur. Monte Carlo simulations often use these high activation barriers, but also naturally consider the reaction-diffusion competition in their kinematic models. The reaction-diffusion competition makes the reactions much more likely to occur. Because we did not incorporate the reaction-diffusion competition, we used an `effective' activation barrier to account for this process. We adopted `effective' barriers of 600\,K for $\rm CO+H$ and 400\,K for $\rm H_2CO+H$ \citep{2005ApJ...626..262A, 2009A&A...505..629F} instead of the high values considered for quantum tunneling when the reaction-diffusion competition is taken into account \citep[$\rm 5400\,K - 9600\,K, $][]{2005ApJ...626..262A}. However, recent experiments by Minissale et al. (priv. comm.) tend to find barriers on the order of 1000\,K. The barriers we used in that case would overestimate the hydrogenation of CO. The exact content of the ices at long time scales will be re-addressed in a future work, once the barriers for the hydrogenation of CO and H2CO are more clearly defined for the solid state.

Higher resolution simulations will allow us to probe the inner pc of a clump, which was not the prime target of this work. We resolved the Jeans length by 44 cells for the initial state, but our Jeans resolution drops to 1.2 cells at the final stages of our simulation (taking into account the increase in mean mololecular mass). This is lower by a factor three than the Truelove criterion \citep{1997ApJ...489L.179T}, but artificial fragmentation at this stage is not a concern, because our simulation concludes shortly after we drop below 4 cells. The Jeans resolution at the final stages of our simulation should not affect the results on which we focused, that is, the chemical reactions.

Higher resolution will also enable obtaining more clumps with a wider range of clump densities and opacities. A statistical study can be performed with such a set of clumps to determine wether our results hold true for the general case. We note that although the chemistry calculations are a local phenomenon, they reflect the cloud history and thus are affected by the evolution of the cloud, which makes the initial conditions (initial abundances, density profile, temperatures, background radiation field) and the large-scale physics (gravity, turbulence, feedback, magnetic fields) important for the results. However, because of its nature, grain surface chemistry acts as a balancing factor to the thermodynamics of the gas by the processes of freeze-out and by its catalytic nature, which causes it to form complex molecules. This, in the end, self-regulate the evolution of the cloud to some extent. To which degree this is true can be realized by a parameter study of simulations with a set of different initial and physical conditions.

A problem arising from our non-conserving numerical solver is that the ion-electron ratio is not conserved. We find that his occurs when the differential equation becomes stiff at higher, $n_{\rm H} \ge 4\times10^3$ \cmcube, densities. The electron abundance diverges after 17 Myrs and is too high. This problem will have negligible effects on the thermal balance, while for the chemical evolution of the species other than electrons, we believe that the remaining simulation time is too short to have a strong impact.

%observational part

\begin{acknowledgements}
This work is supported by the Netherlands Organization for Scientific Research (NWO). The authors thank P. Caselli, M. Spaans, M. Minissale, and F. Dulieu for all the useful discussions on various aspects of this work. SH thanks Malcolm Walmsley for pointing out the error in the ion-balance. The software used in this work was developed in part by the DOE NNSA ASC- and DOE Office of Science ASCR-supported Flash Center for Computational Science at the University of Chicago. The simulations have been run on the dedicated special purpose machines at the Kapteyn Astronomical Institute and at the Donald Smits Center for Information Technology using the Millipede Cluster, University of Groningen. Some of the kinetic data implemented in this work have been downloaded from the online database \textit{KiDA} (\cite{2012ApJS..199...21W}, http://kida.obs.u-bordeaux1.fr). 
\end{acknowledgements}

%\bibliography{biblio.seventh.bib}

\appendix
%========
\section{Reaction tables}
Here we list in tables the gas-phase and the surface reactions. Since adsorption reactions (Sect. \ref{sec:accretion}) and thermal desorption reactions (Sect. \ref{sec:evaporation}) are relatively straightforward, they are omitted from this list. We also supply the adopted activation barriers, E$_a$, in Kelvin within the relevant tables.

\begin{table}[htb!]
\begin{small}
\caption{Photo-processes on surfaces.}
\begin{tabular}{lll}
\hline
\hline
Reactants                       &               & Products                      \\
\hline
$\bot$H$_2$ + CR                & $\rightarrow$ & $\bot$H + $\bot$H                 \\
$\bot$O$_2$ + CRUV              & $\rightarrow$ & $\bot$O + $\bot$O                 \\
$\bot$OH + CRUV                 & $\rightarrow$ & $\bot$H + $\bot$O                 \\
$\bot$H$_2$O + CRUV             & $\rightarrow$ & $\bot$H + $\bot$OH                 \\
$\bot$CO$_2$ + CRUV             & $\rightarrow$ & $\bot$O + $\bot$CO                 \\
$\bot$HCO + CRUV                & $\rightarrow$ & $\bot$H + $\bot$CO                 \\
$\bot$H$_2$CO + CRUV            & $\rightarrow$ & $\bot$H + $\bot$HCO                 \\
$\bot$CH$_3$O + CRUV            & $\rightarrow$ & $\bot$H + $\bot$H$_2$CO         \\
$\bot$CH$_3$OH + CRUV           & $\rightarrow$ & $\bot$H + $\bot$CH$_3$O         \\
$\bot$H$_2$ + UV Photon         & $\rightarrow$ & $\bot$H + $\bot$H                 \\
$\bot$O$_2$ + UV Photon         & $\rightarrow$ & $\bot$O + $\bot$O                 \\
$\bot$OH + UV Photon            & $\rightarrow$ & $\bot$H + $\bot$O                 \\
$\bot$H$_2$O + UV Photon        & $\rightarrow$ & $\bot$H + $\bot$OH                 \\
$\bot$CO$_2$ + UV Photon        & $\rightarrow$ & $\bot$O + $\bot$CO                 \\
$\bot$HCO + UV Photon           & $\rightarrow$ & $\bot$H + $\bot$CO                 \\
$\bot$H$_2$CO + UV Photon       & $\rightarrow$ & $\bot$H + $\bot$HCO                 \\
$\bot$CH$_3$O + UV Photon       & $\rightarrow$ & $\bot$H + $\bot$H$_2$CO         \\
$\bot$CH$_3$OH + UV Photon      & $\rightarrow$ & $\bot$H + $\bot$CH$_3$O         \\
$\bot$CO + UV Photon            & $\rightarrow$ & CO                            \\
$\bot$CO + CRUV                 & $\rightarrow$ & CO                            \\
\hline
\end{tabular}
\label{tab:appendix4}
\end{small}
\end{table}

\begin{table*}[htb!]
\begin{small}
\caption{Photo-processes in the gas phase.}
\begin{tabular}{lll||lll||lll}
\hline
\hline
Reactants & & Products & Reactants & & Products & Reactants & & Products \\
\hline
 H         + CR        & $\rightarrow$ & H$^+$        + e$^-$        &
 O         + CR        & $\rightarrow$ & O$^+$        + e$^-$        &
 CO        + CR        & $\rightarrow$ & C         + O         \\
 H$_2$        + CR        & $\rightarrow$ & H         + H         &
 H$_2$        + CR        & $\rightarrow$ & H         + H$^+$        + e$^-$        &
 H$_2$        + CR        & $\rightarrow$ & H$^+$        + H$^-$        \\
 H$_2$        + CR        & $\rightarrow$ & H$_2$$^+$       + e$^-$        &
 O$_2$        + CRUV       & $\rightarrow$ & O         + O         &
 OH        + CRUV       & $\rightarrow$ & H         + O         \\
 CO$_2$       + CRUV       & $\rightarrow$ & O         + CO        &
 H$_2$O       + CRUV       & $\rightarrow$ & H         + OH        &
 HCO       + CRUV       & $\rightarrow$ & H         + CO        \\
 HCO       + CRUV       & $\rightarrow$ & HCO$^+$      + e$^-$        &
 H$_2$CO      + CRUV       & $\rightarrow$ & CO        + H$_2$        &
 CH$_4$OH      + CRUV       & $\rightarrow$ & H$_2$        + H$_2$CO      \\
 C         + CRUV       & $\rightarrow$ & C$^+$        + e$^-$        &
 CH$_4$OH      + Photon    & $\rightarrow$ & H$_2$        + H$_2$CO      &
 H$_2$$^+$       + Photon    & $\rightarrow$ & H         + H$^+$        \\
 OH$^+$       + Photon    & $\rightarrow$ & O         + H$^+$        &
 H$_3$$^+$       + Photon    & $\rightarrow$ & H$_2$        + H$^+$        &
 H$_3$$^+$       + Photon    & $\rightarrow$ & H         + H$_2$$^+$       \\
 C$^-$        + Photon    & $\rightarrow$ & C         + e$^-$        &
 H$^-$        + Photon    & $\rightarrow$ & H         + e$^-$        &
 O$^-$        + Photon    & $\rightarrow$ & O         + e$^-$        \\
 C         + Photon    & $\rightarrow$ & C$^+$        + e$^-$        &
 CO        + Photon    & $\rightarrow$ & C         + O         &
 H$_2$        + Photon    & $\rightarrow$ & H         + H         \\
 O$_2$        + Photon    & $\rightarrow$ & O         + O         &
 OH        + Photon    & $\rightarrow$ & H         + O         &
 OH        + Photon    & $\rightarrow$ & OH$^+$       + e$^-$        \\
 CO$_2$       + Photon    & $\rightarrow$ & O         + CO        &
 H$_2$O       + Photon    & $\rightarrow$ & H         + OH        &
 H$_2$O       + Photon    & $\rightarrow$ & H$_2$O$^+$      + e$^-$        \\
 HCO       + Photon    & $\rightarrow$ & H         + CO        &
 HCO       + Photon    & $\rightarrow$ & HCO$^+$      + e$^-$        &
 H$_2$CO      + Photon    & $\rightarrow$ & H         + H         + CO        \\
 H$_2$CO      + Photon    & $\rightarrow$ & CO        + H$_2$        &
 H$_2$CO      + Photon    & $\rightarrow$ & H         + HCO$^+$      + e$^-$       \\
\hline
\end{tabular}
\label{tab:appendixG1}
\end{small}
\end{table*}

\begin{table*}[htb!]
\begin{small}
\caption{Gas-phase reactions.}
\begin{tabular}{lll||lll||lll}
\hline
\hline
Reactants & & Products & Reactants & & Products & Reactants & & Products \\
\hline
 H$_2$O       + C$^+$        & $\rightarrow$ & H         + HCO$^+$      &
 O$_2$        + C$^-$        & $\rightarrow$ & CO        + O$^-$        &
 CO$_2$       + C$^-$        & $\rightarrow$ & CO        + CO        + e$^-$        \\
 O         + HCO       & $\rightarrow$ & H         + CO$_2$       &
 O         + HCO       & $\rightarrow$ & CO        + OH        &
 CO        + OH        & $\rightarrow$ & H         + CO$_2$       \\
 OH        + OH        & $\rightarrow$ & O         + H$_2$O       &
 OH        + OH        & $\rightarrow$ & O         + H$_2$O       &
 OH        + HCO       & $\rightarrow$ & CO        + H$_2$O       \\
 OH        + H$_2$CO      & $\rightarrow$ & H$_2$O       + HCO       &
 O         + H$_2$$^+$       & $\rightarrow$ & H         + OH$^+$       &
 CO        + H$_2$$^+$       & $\rightarrow$ & H         + HCO$^+$      \\
 H$_2$        + H$_2$$^+$       & $\rightarrow$ & H         + H$_3$$^+$       &
 OH        + H$_2$$^+$       & $\rightarrow$ & H         + H$_2$O$^+$      &
 H$_2$O       + H$_2$$^+$       & $\rightarrow$ & H         + H$_3$O$^+$      \\
 HCO       + H$_2$$^+$       & $\rightarrow$ & CO        + H$_3$$^+$       &
 H$_2$CO      + H$_2$$^+$       & $\rightarrow$ & H         + H$_2$        + HCO$^+$      &
 H$_3$$^+$       + H$^-$        & $\rightarrow$ & H$_2$        + H$_2$        \\
 HCO$^+$      + H$^-$        & $\rightarrow$ & CO        + H$_2$        &
 H$_3$O$^+$      + H$^-$        & $\rightarrow$ & H         + H$_2$        + OH        &
 H$_3$O$^+$      + H$^-$        & $\rightarrow$ & H$_2$        + H$_2$O       \\
 OH        + HCO$^+$      & $\rightarrow$ & CO        + H$_2$O$^+$      &
 H$_2$O       + HCO$^+$      & $\rightarrow$ & CO        + H$_3$O$^+$      &
 HCO       + HCO       & $\rightarrow$ & CO        + H$_2$CO      \\
 H         + HCO       & $\rightarrow$ & CO        + H$_2$        &
 H         + OH        & $\rightarrow$ & O         + H$_2$        &
 H         + H$_2$O       & $\rightarrow$ & H$_2$        + OH        \\
 H         + H$_2$CO      & $\rightarrow$ & H$_2$        + HCO       &
 H         + O$_2$        & $\rightarrow$ & O         + OH        &
 H         + CO$_2$       & $\rightarrow$ & CO        + OH        \\
 O         + H$_2$        & $\rightarrow$ & H         + OH        &
 H$_2$        + OH        & $\rightarrow$ & H         + H$_2$O       &
 O         + OH        & $\rightarrow$ & H         + O$_2$        \\
 O         + OH        & $\rightarrow$ & H         + O$_2$        &
 O         + OH        & $\rightarrow$ & H         + O$_2$        &
 CO        + H$_2$O$^+$      & $\rightarrow$ & OH        + HCO$^+$      \\
 H$_2$        + H$_2$O$^+$      & $\rightarrow$ & H         + H$_3$O$^+$      &
 OH        + H$_2$O$^+$      & $\rightarrow$ & O         + H$_3$O$^+$      &
 H$_2$O       + H$_2$O$^+$      & $\rightarrow$ & OH        + H$_3$O$^+$      \\
 HCO       + H$_2$O$^+$      & $\rightarrow$ & CO        + H$_3$O$^+$      &
 O         + H$_3$$^+$       & $\rightarrow$ & H$_2$        + OH$^+$       &
 O         + H$_3$$^+$       & $\rightarrow$ & H         + H$_2$O$^+$      \\
 CO        + H$_3$$^+$       & $\rightarrow$ & H$_2$        + HCO$^+$      &
 OH        + H$_3$$^+$       & $\rightarrow$ & H$_2$        + H$_2$O$^+$      &
 H$_2$O       + H$_3$$^+$       & $\rightarrow$ & H$_2$        + H$_3$O$^+$      \\
 CO$_2$       + H$^+$        & $\rightarrow$ & O         + HCO$^+$      &
 HCO       + H$^+$        & $\rightarrow$ & CO        + H$_2$$^+$       &
 H$_2$        + O$^+$        & $\rightarrow$ & H         + OH$^+$       \\
 HCO       + O$^+$        & $\rightarrow$ & CO        + OH$^+$       &
 H$_2$CO      + O$^+$        & $\rightarrow$ & OH        + HCO$^+$      &
 C         + H$_3$O$^+$      & $\rightarrow$ & H$_2$        + HCO$^+$      \\
 H$_2$CO      + H$^+$        & $\rightarrow$ & H$_2$        + HCO$^+$      &
 CO        + OH$^+$       & $\rightarrow$ & O         + HCO$^+$      &
 H$_2$        + OH$^+$       & $\rightarrow$ & H         + H$_2$O$^+$      \\
 OH        + OH$^+$       & $\rightarrow$ & O         + H$_2$O$^+$      &
 H$_2$O       + OH$^+$       & $\rightarrow$ & O         + H$_3$O$^+$      &
 HCO       + OH$^+$       & $\rightarrow$ & CO        + H$_2$O$^+$      \\
 C         + O$_2$        & $\rightarrow$ & O         + CO        &
 C         + OH        & $\rightarrow$ & H         + CO        &
 HCO$^+$      + C$^-$        & $\rightarrow$ & C         + H         + CO        \\
 H$_3$O$^+$      + C$^-$        & $\rightarrow$ & C         + H         + H$_2$O       &
 H$_3$$^+$       + C$^-$        & $\rightarrow$ & C         + H         + H$_2$        &
 HCO$^+$      + O$^-$        & $\rightarrow$ & H         + O         + CO        \\
 H$_3$O$^+$      + O$^-$        & $\rightarrow$ & H         + O         + H$_2$O       &
 H$_3$$^+$       + O$^-$        & $\rightarrow$ & H         + O         + H$_2$        &
 H         + H$_2$$^+$       & $\rightarrow$ & H$_2$        + H$^+$        \\
 C$^+$        + C$^-$        & $\rightarrow$ & C         + C         &
 C$^+$        + H$^-$        & $\rightarrow$ & C         + H         &
 H$^+$        + C$^-$        & $\rightarrow$ & C         + H         \\
 H$^+$        + H$^-$        & $\rightarrow$ & H         + H         &
 O$^+$        + C$^-$        & $\rightarrow$ & C         + O         &
 O$^+$        + H$^-$        & $\rightarrow$ & H         + O         \\
 H$_2$$^+$       + H$^-$        & $\rightarrow$ & H         + H$_2$        &
 O         + H$^+$        & $\rightarrow$ & H         + O$^+$        &
 OH        + H$^+$        & $\rightarrow$ & H         + OH$^+$       \\
 H$_2$O       + H$^+$        & $\rightarrow$ & H         + H$_2$O$^+$      &
 H         + O$^+$        & $\rightarrow$ & O         + H$^+$        &
 H$_2$O       + O$^+$        & $\rightarrow$ & O         + H$_2$O$^+$      \\
 HCO       + C$^+$        & $\rightarrow$ & C         + HCO$^+$      &
 HCO       + O$^+$        & $\rightarrow$ & O         + HCO$^+$      &
 OH        + O$^+$        & $\rightarrow$ & O         + OH$^+$       \\
 OH        + H$_2$$^+$       & $\rightarrow$ & H$_2$        + OH$^+$       &
 H$_2$O       + H$_2$$^+$       & $\rightarrow$ & H$_2$        + H$_2$O$^+$      &
 HCO       + H$_2$$^+$       & $\rightarrow$ & H$_2$        + HCO$^+$      \\
 HCO       + H$_2$O$^+$      & $\rightarrow$ & H$_2$O       + HCO$^+$      &
 HCO       + H$^+$        & $\rightarrow$ & H         + HCO$^+$      &
 H$_2$O       + OH$^+$       & $\rightarrow$ & OH        + H$_2$O$^+$      \\
 HCO       + OH$^+$       & $\rightarrow$ & OH        + HCO$^+$      &
 HCO$^+$      + C$^-$        & $\rightarrow$ & C         + HCO       &
 HCO$^+$      + O$^-$        & $\rightarrow$ & O         + HCO       \\
 H$^+$        + O$^-$        & $\rightarrow$ & H         + O         &
 H         + H$^+$        & $\rightarrow$ & H$_2$$^+$       + Photon    &
 C         + O         & $\rightarrow$ & CO        + Photon    \\
 H         + O         & $\rightarrow$ & OH        + Photon    &
 O         + O         & $\rightarrow$ & O$_2$        + Photon    &
 H         + OH        & $\rightarrow$ & H$_2$O       + Photon    \\
 O         + C$^-$        & $\rightarrow$ & CO        + e$^-$        &
 O$_2$        + C$^-$        & $\rightarrow$ & CO$_2$       + e$^-$        &
 OH        + C$^-$        & $\rightarrow$ & HCO       + e$^-$        \\
 H$_2$O       + C$^-$        & $\rightarrow$ & H$_2$CO      + e$^-$        &
 H         + H$^-$        & $\rightarrow$ & H$_2$        + e$^-$        &
 O         + H$^-$        & $\rightarrow$ & OH        + e$^-$        \\
 CO        + H$^-$        & $\rightarrow$ & HCO       + e$^-$        &
 OH        + H$^-$        & $\rightarrow$ & H$_2$O       + e$^-$        &
 HCO       + H$^-$        & $\rightarrow$ & H$_2$CO      + e$^-$        \\
 C         + O$^-$        & $\rightarrow$ & CO        + e$^-$        &
 H         + O$^-$        & $\rightarrow$ & OH        + e$^-$        &
 O         + O$^-$        & $\rightarrow$ & O$_2$        + e$^-$        \\
 CO        + O$^-$        & $\rightarrow$ & CO$_2$       + e$^-$        &
 H$_2$        + O$^-$        & $\rightarrow$ & H$_2$O       + e$^-$        &
 C         + e$^-$        & $\rightarrow$ & C$^-$                   \\
 H         + e$^-$        & $\rightarrow$ & H$^-$                   &
 O         + e$^-$        & $\rightarrow$ & O$^-$                   &
 H$_2$$^+$       + e$^-$        & $\rightarrow$ & H         + H         \\
 H$_2$$^+$       + e$^-$        & $\rightarrow$ & H$_2$        + Photon    &
 OH$^+$       + e$^-$        & $\rightarrow$ & H         + O         &
 H$_2$O$^+$      + e$^-$        & $\rightarrow$ & O         + H$_2$        \\
 H$_2$O$^+$      + e$^-$        & $\rightarrow$ & H         + OH        &
 H$_2$O$^+$      + e$^-$        & $\rightarrow$ & H         + H         + O         &
 H$_3$$^+$       + e$^-$        & $\rightarrow$ & H         + H         + H         \\
 H$_3$$^+$       + e$^-$        & $\rightarrow$ & H         + H$_2$        &
 HCO$^+$      + e$^-$        & $\rightarrow$ & H         + CO        &
 C$^+$        + e$^-$        & $\rightarrow$ & C         + Photon    \\
 H$^+$        + e$^-$        & $\rightarrow$ & H         + Photon    &
 O$^+$        + e$^-$        & $\rightarrow$ & O         + Photon    &
 H$_3$O$^+$      + e$^-$        & $\rightarrow$ & H         + H         + OH        \\
 H$_3$O$^+$      + e$^-$        & $\rightarrow$ & H         + H$_2$O       &
 H$_3$O$^+$      + e$^-$        & $\rightarrow$ & H$_2$        + OH        &
 H$_3$O$^+$      + e$^-$        & $\rightarrow$ & H         + O         + H$_2$ \\
 O$_2$        + C$^+$        & $\rightarrow$ & CO        + O$^+$        \\
\hline
\end{tabular}
\label{tab:appendixG2}
\end{small}
\end{table*}

\begin{table}[htb!]
\begin{small}
\caption{Surface reactions leading to surface products.}
\begin{tabular}{lll|c|c}
\hline
\hline
Reactants                       &               & Products                      & E$_a$/K & Reference \\
\hline
$\bot$H + $\bot$H               & $\rightarrow$ & $\bot$H$_2$                         & 0             & \\
$\bot$H + $\bot$O               & $\rightarrow$ & $\bot$OH                      & 0               & \\
$\bot$H + $\bot$OH              & $\rightarrow$ & $\bot$H$_2$O                         & 0             & \\
$\bot$H + $\bot$O$_3$           & $\rightarrow$ & $\bot$OH + $\bot$O$_2$     & 480           & c \\
$\bot$H + $\bot$O$_3$           & $\rightarrow$ & $\bot$OH + $\bot$O$_2$     &               & i,a(450) \\
$\bot$H + $\bot$H$_2$O$_2$      & $\rightarrow$ & $\bot$OH + $\bot$H$_2$O    & 1000          & o(800-1250) \\
$\bot$H + $\bot$O$_2$           & $\rightarrow$ & $\bot$HO$_2$                         & 200           & b(0-250) \\
$\bot$H + $\bot$O$_2$           & $\rightarrow$ & $\bot$HO$_2$                         &               & j(0-200) \\
$\bot$H + $\bot$H$_2$O          & $\rightarrow$ & $\bot$OH + $\bot$H$_2$     & 9600          & r \\
$\bot$H + $\bot$HO$_2$          & $\rightarrow$ & $\bot$OH + $\bot$OH                & 0             & \\
$\bot$H + $\bot$CO              & $\rightarrow$ & $\bot$HCO                     & 600             & h, k \\ %(300-770) \\%, k(390-520) \\
%                               &               &                       &                                               & k(390-520) \\
$\bot$H + $\bot$HCO             & $\rightarrow$ & $\bot$H$_2$CO                 & 0             & \\
$\bot$H + $\bot$H$_2$CO         & $\rightarrow$ & $\bot$CH$_3$O                 & 400           & h, k \\ %(330-800) \\%, k(415-490) \\
%                               &               &                               &                                       & k(415-490) \\
$\bot$H + $\bot$CH$_3$O         & $\rightarrow$ & $\bot$CH$_3$OH                 & 0             & \\
$\bot$H + $\bot$HCO             & $\rightarrow$ & $\bot$CO + $\bot$H$_2$     & 400           & q \\
$\bot$H + $\bot$H$_2$CO         & $\rightarrow$ & $\bot$HCO $\bot$H$_2$         & 2250          & n \\
$\bot$H + $\bot$CH$_3$O         & $\rightarrow$ & $\bot$H$_2$CO + $\bot$H$_2$   & 150           & n \\
$\bot$H + $\bot$CH$_3$OH        & $\rightarrow$ & $\bot$CH$_3$O + $\bot$H$_2$   & 3000          & n \\
$\bot$H + $\bot$CO$_2$          & $\rightarrow$ & $\bot$CO + $\bot$OH                & 10000         & n \\
$\bot$O + $\bot$O               & $\rightarrow$ & $\bot$O$_2$                         & 0             & \\
$\bot$O + $\bot$O$_2$           & $\rightarrow$ & $\bot$O$_3$                         & 0             & \\
$\bot$O + $\bot$O$_3$           & $\rightarrow$ & $\bot$O$_2$ + $\bot$O$_2$   & 2300          & p \\
$\bot$O + $\bot$O$_3$           & $\rightarrow$ & $\bot$O$_2$ + $\bot$O$_2$   &               & r(2000) \\
$\bot$O + $\bot$HO$_2$          & $\rightarrow$ & $\bot$O$_2$ + $\bot$OH      & 0             & \\
$\bot$O + $\bot$H$_2$           & $\rightarrow$ & $\bot$H + $\bot$OH                 & 4640          & l \\%, o(3165) \\
$\bot$O + $\bot$OH              & $\rightarrow$ & $\bot$O$_2$ + $\bot$H               & 0             & \\
$\bot$O + $\bot$CO              & $\rightarrow$ & $\bot$CO$_2$                         & 160           & p($\geq$160) \\
$\bot$O + $\bot$CO              & $\rightarrow$ & $\bot$CO$_2$                         &               & e(290) \\
$\bot$O + $\bot$HCO             & $\rightarrow$ & $\bot$CO$_2$ + $\bot$H       & 0             & \\
$\bot$O + $\bot$H$_2$CO         & $\rightarrow$ & $\bot$CO$_2$ + $\bot$H$_2$   & 300           & q \\
$\bot$OH + $\bot$OH             & $\rightarrow$ & $\bot$H$_2$O$_2$                 & 0             & \\
$\bot$OH + $\bot$H$_2$          & $\rightarrow$ & $\bot$H + $\bot$H$_2$O         & 2100          & g \\
$\bot$OH + $\bot$CO             & $\rightarrow$ & $\bot$CO$_2$ + $\bot$H       & 600           & d \\%, d(554)
$\bot$OH + $\bot$CO             & $\rightarrow$ & $\bot$CO$_2$ + $\bot$H       &               & f(519),m(400) \\%, d(554)
$\bot$OH + $\bot$HCO            & $\rightarrow$ & $\bot$CO$_2$ + $\bot$H$_2$   & 0             & \\
$\bot$HO$_2$ + $\bot$H$_2$      & $\rightarrow$ & $\bot$H + $\bot$H$_2$O$_2$         & 5000          & o \\
\hline
\end{tabular}
\label{tab:appendix1} \\
Note 1: The parentheses give barriers from respective studies. \\
Note 2: Extra references given for reactions with similar barriers. \\
Note 3: For unknown/uncertain barriers, the \textit{NIST}$^{n}$ database is used. \\
Note 4: The barriers for $\bot$H+$\bot$CO and $\bot$H+$\bot$H$_2$CO are `effective' \\
barriers deduced from {\it h,k} using their rates. \\
$^{a}$ \cite{1978JChPh..69..350L}, \\
$^{b}$ \cite{1988JChPh..88.6273W}, \\
$^{c}$ \cite{1989JPCRD..18..881A}, \\
$^{d}$ \cite{2000JChPh.113.5138D}, \\
$^{e}$ \cite{2001ApJ...555L..61R}, \\
$^{f}$ \cite{2001CPL...349..547Y}, \\
$^{g}$ \cite{2004ACP.....4.1461A}, \\
$^{h}$ \cite{2005ApJ...626..262A}, \\
$^{i}$ \cite{2007ApJ...668..294C}, \\
$^{j}$ \cite{2008CPL...456...27M}, \\
$^{k}$ \cite{2009A&A...505..629F}, \\
$^{l}$ \cite{2010ApJ...713..662A}, \\
$^{m}$ \cite{2011ApJ...735..121N}, \\
$^{n}$ \textit{NIST} database; http://kinetics.nist.gov \\ %\cite{NIST}, \\
$^{o}$ \cite{2013PCCP...15.8287L}, \\
$^{p}$ \cite{2013A&A...559A..49M}, \\
$^{q}$ Best estimate (priv. comm. Minissale), \\ %Minissale et al. (2014) in prep. \\
$^{r}$ \cite{2010A&A...522A..74C}
\end{small}
\end{table}

\begin{table}[htb!]
\begin{small}
\caption{Chemical desorption reactions.}
\begin{tabular}{lll|c|c}
\hline
\hline
Reactants                       &               & Products              & E$_a$/K         & Reference \\
\hline
$\bot$H + $\bot$H               & $\rightarrow$ & H$_2$                 & 0               & \\
$\bot$H + $\bot$O               & $\rightarrow$ & OH                    & 0               & \\
$\bot$H + $\bot$OH              & $\rightarrow$ & H$_2$O                & 0               & \\
$\bot$H + $\bot$O$_3$           & $\rightarrow$ & OH + O$_2$                 & 480           & See Tab. \ref{tab:appendix1} \\
$\bot$H + $\bot$H$_2$O$_2$      & $\rightarrow$ & OH + H$_2$O                 & 1000          & See Tab. \ref{tab:appendix1} \\
$\bot$H + $\bot$CO              & $\rightarrow$ & HCO                   & 600             & See Tab. \ref{tab:appendix1} \\
$\bot$H + $\bot$HCO             & $\rightarrow$ & H$_2$CO               & 0               & \\
$\bot$H + $\bot$H$_2$CO         & $\rightarrow$ & CH$_3$O               & 400             & See Tab. \ref{tab:appendix1} \\
$\bot$H + $\bot$CH$_3$O         & $\rightarrow$ & CH$_3$OH              & 0               & \\
$\bot$O + $\bot$O               & $\rightarrow$ & O$_2$                 & 0               & \\
$\bot$O + $\bot$O$_3$           & $\rightarrow$ & O$_2$ + O$_2$         & 0 & \\ %added by mistake, but makes no difference
$\bot$O + $\bot$HO$_2$          & $\rightarrow$ & O$_2$ + OH                 & 0 & \\ %added by mistake, but makes no difference
\hline
\end{tabular}
\label{tab:appendix2}
\end{small}
\end{table}

\begin{table}[htb!]
\begin{small}
\caption{Chemisorption reactions.}
\begin{tabular}{lll|c}
\hline
\hline
Reactants               &               & Products      & E$_a$/K \\
\hline
~~~H                    & $\rightarrow$ & $\bot$H$_c$   & 1000 \\
~~~H + $\bot$H$_c$      & $\rightarrow$ & ~~~H$_2$      & 1000 \\
$\bot$H                 & $\rightarrow$ & $\bot$H$_c$   & 10000 \\
$\bot$H$_c$             & $\rightarrow$ & ~~~H          & see \cite{2010ApJ...715..698C} \\
$\bot$H + $\bot$H$_c$   & $\rightarrow$ & ~~~H$_2$      & see \cite{2010ApJ...715..698C} \\
\hline
\end{tabular}
\label{tab:appendix3}
\end{small}
\end{table}

\bibliography{biblio.seventh.bib} %ARXIV

\end{document}